\documentclass[aps,prd,showpacs,twocolumn,superscriptaddress,floatfix]{revtex4}  

\usepackage{rotating}
\usepackage{subfigure}
\usepackage{graphicx}
\usepackage{dcolumn}
\usepackage{bm}
\usepackage{amssymb} 

\usepackage{epsfig}
\usepackage{psfrag}
\usepackage{appendix}
\usepackage{enumerate}
\usepackage{multirow}

\usepackage{color}

\newcommand{\comment}[1]{} %

\newcommand{\MET}{{\mbox{$E\kern-0.57em\raise0.19ex\hbox{/}_{T}$}}}
\newcommand{\met}{{\mbox{$E\kern-0.57em\raise0.19ex\hbox{/}_{T}$}}}

\newcommand{\ifb}{fb$^{-1}$}
\newcommand{\pp}{$p\bar{p}$}

\newcommand{\ttbar}{$t\bar{t}$}

\newcommand{\pvh}{$q\bar{q}\rightarrow VH$}
\newcommand{\pvbf}{$q\bar{q} \rightarrow q^{\prime}\bar{q}^{\prime} H_{f}$}
\newcommand{\pggh}{$gg\rightarrow H$}

\newcommand{\htt}{$H\rightarrow \tau^+ \tau^-$}

\newcommand{\whl}{$WH\rightarrow \ell\nu b\bar{b}$}

\newcommand{\zhl}{$ZH\rightarrow \ell\ell b\bar{b}$}

\newcommand{\zhv}{$ZH\rightarrow \nu\bar{\nu} b\bar{b}$}

\newcommand{\vhvww}{$VH \rightarrow VWW$}
\newcommand{\ssem}{$VH \rightarrow e^\pm \mu^\pm$+$X$}
\newcommand{\trilep}{$VH \rightarrow ee\mu/\mu\mu e$+$X$}
\newcommand{\ttm}{$VH \rightarrow \tau_{h}\tau_{h}\mu+X$}
\newcommand{\tautaujj}{$H+X \rightarrow \ell \tau_{h} jj$}

\newcommand{\hww}{$H\rightarrow W^+ W^-$}

\newcommand{\hwweemm}{$H\rightarrow W^+ W^-\rightarrow e\nu e\nu /\mu\nu\mu\nu$}
\newcommand{\hwwlnulnu}{$H\rightarrow W^+ W^-\rightarrow (e^+e^-,\mu^+\mu^-,e^{\pm}\mu^{\mp}) \nu \bar{\nu}$}
\newcommand{\hwwmvtv}{$H+X\rightarrow W^+ W^-\rightarrow \mu^\pm \tau^{\mp}_{h}+\leq 1~\text{jet}$}
\newcommand{\hwwlnuqq}{$H\rightarrow W^+ W^-\rightarrow\ell \nu q^{\prime}\bar{q}$}
\newcommand{\lnuqqqq}{$VH\rightarrow \ell \nu q^{\prime}\bar{q}q^{\prime}\bar{q}$}
\newcommand{\hwwlvlv}{$H\rightarrow W^+ W^-\rightarrow \ell^{+}\nu\ell^{-}\bar{\nu}$}

\newcommand{\vbf}{$q\bar{q} \rightarrow q^{\prime}\bar{q}^{\prime} H$}
\newcommand{\hgg}{$H\rightarrow \gamma \gamma$}
\newcommand{\hbb}{$H\rightarrow b\bar{b}$}

\newcommand{\sigmaggh}{$\sigma_{gg\to H}$}

\newcommand{\tevE}{$\sqrt{s}=1.96$~TeV}
\newcommand{\gev}{~Ge\kern -0.05em V\kern -0.1em /$c^2$}
\newcommand{\dgev}{Ge\kern -0.05em V\kern -0.1em /$c^2$}

\newcommand{\lumimax}{9.7}
\newcommand{\obsABE}{2.86} 
\newcommand{\obsAFE}{0.66} 
\newcommand{\expABE}{1.68} 
\newcommand{\expAFE}{0.70} 
\newcommand{\exclmin}{157} 
\newcommand{\exclmax}{178}
\newcommand{\excllow}{101} 
\newcommand{\exclminexp}{155}
\newcommand{\exclmaxexp}{175} 
\newcommand{\ADZlocalpval}{1.8\%} 
\newcommand{\ADZlocalzval}{2.1} 
\newcommand{\ABElocalpval}{4.1\%} 
\newcommand{\ABElocalzval}{1.7}
\newcommand\T{\rule{0pt}{2.6ex}}       

\begin{document}

\hspace{5in}\mbox{FERMILAB-PUB-13-059-E}

\vspace*{1.0cm}

\title{Combined search for the Higgs boson with the D0 experiment}

\affiliation{LAFEX, Centro Brasileiro de Pesquisas F\'{i}sicas, Rio de Janeiro, Brazil}
\affiliation{Universidade do Estado do Rio de Janeiro, Rio de Janeiro, Brazil}
\affiliation{Universidade Federal do ABC, Santo Andr\'e, Brazil}
\affiliation{University of Science and Technology of China, Hefei, People's Republic of China}
\affiliation{Universidad de los Andes, Bogot\'a, Colombia}
\affiliation{Charles University, Faculty of Mathematics and Physics, Center for Particle Physics, Prague, Czech Republic}
\affiliation{Czech Technical University in Prague, Prague, Czech Republic}
\affiliation{Center for Particle Physics, Institute of Physics, Academy of Sciences of the Czech Republic, Prague, Czech Republic}
\affiliation{Universidad San Francisco de Quito, Quito, Ecuador}
\affiliation{LPC, Universit\'e Blaise Pascal, CNRS/IN2P3, Clermont, France}
\affiliation{LPSC, Universit\'e Joseph Fourier Grenoble 1, CNRS/IN2P3, Institut National Polytechnique de Grenoble, Grenoble, France}
\affiliation{CPPM, Aix-Marseille Universit\'e, CNRS/IN2P3, Marseille, France}
\affiliation{LAL, Universit\'e Paris-Sud, CNRS/IN2P3, Orsay, France}
\affiliation{LPNHE, Universit\'es Paris VI and VII, CNRS/IN2P3, Paris, France}
\affiliation{CEA, Irfu, SPP, Saclay, France}
\affiliation{IPHC, Universit\'e de Strasbourg, CNRS/IN2P3, Strasbourg, France}
\affiliation{IPNL, Universit\'e Lyon 1, CNRS/IN2P3, Villeurbanne, France and Universit\'e de Lyon, Lyon, France}
\affiliation{III. Physikalisches Institut A, RWTH Aachen University, Aachen, Germany}
\affiliation{Physikalisches Institut, Universit\"at Freiburg, Freiburg, Germany}
\affiliation{II. Physikalisches Institut, Georg-August-Universit\"at G\"ottingen, G\"ottingen, Germany}
\affiliation{Institut f\"ur Physik, Universit\"at Mainz, Mainz, Germany}
\affiliation{Ludwig-Maximilians-Universit\"at M\"unchen, M\"unchen, Germany}
\affiliation{Fachbereich Physik, Bergische Universit\"at Wuppertal, Wuppertal, Germany}
\affiliation{Panjab University, Chandigarh, India}
\affiliation{Delhi University, Delhi, India}
\affiliation{Tata Institute of Fundamental Research, Mumbai, India}
\affiliation{University College Dublin, Dublin, Ireland}
\affiliation{Korea Detector Laboratory, Korea University, Seoul, Korea}
\affiliation{CINVESTAV, Mexico City, Mexico}
\affiliation{Nikhef, Science Park, Amsterdam, the Netherlands}
\affiliation{Radboud University Nijmegen, Nijmegen, the Netherlands}
\affiliation{Joint Institute for Nuclear Research, Dubna, Russia}
\affiliation{Institute for Theoretical and Experimental Physics, Moscow, Russia}
\affiliation{Moscow State University, Moscow, Russia}
\affiliation{Institute for High Energy Physics, Protvino, Russia}
\affiliation{Petersburg Nuclear Physics Institute, St. Petersburg, Russia}
\affiliation{Instituci\'{o} Catalana de Recerca i Estudis Avan\c{c}ats (ICREA) and Institut de F\'{i}sica d'Altes Energies (IFAE), Barcelona, Spain}
\affiliation{Uppsala University, Uppsala, Sweden}
\affiliation{Lancaster University, Lancaster LA1 4YB, United Kingdom}
\affiliation{Imperial College London, London SW7 2AZ, United Kingdom}
\affiliation{The University of Manchester, Manchester M13 9PL, United Kingdom}
\affiliation{University of Arizona, Tucson, Arizona 85721, USA}
\affiliation{University of California Riverside, Riverside, California 92521, USA}
\affiliation{Florida State University, Tallahassee, Florida 32306, USA}
\affiliation{Fermi National Accelerator Laboratory, Batavia, Illinois 60510, USA}
\affiliation{University of Illinois at Chicago, Chicago, Illinois 60607, USA}
\affiliation{Northern Illinois University, DeKalb, Illinois 60115, USA}
\affiliation{Northwestern University, Evanston, Illinois 60208, USA}
\affiliation{Indiana University, Bloomington, Indiana 47405, USA}
\affiliation{Purdue University Calumet, Hammond, Indiana 46323, USA}
\affiliation{University of Notre Dame, Notre Dame, Indiana 46556, USA}
\affiliation{Iowa State University, Ames, Iowa 50011, USA}
\affiliation{University of Kansas, Lawrence, Kansas 66045, USA}
\affiliation{Louisiana Tech University, Ruston, Louisiana 71272, USA}
\affiliation{Northeastern University, Boston, Massachusetts 02115, USA}
\affiliation{University of Michigan, Ann Arbor, Michigan 48109, USA}
\affiliation{Michigan State University, East Lansing, Michigan 48824, USA}
\affiliation{University of Mississippi, University, Mississippi 38677, USA}
\affiliation{University of Nebraska, Lincoln, Nebraska 68588, USA}
\affiliation{Rutgers University, Piscataway, New Jersey 08855, USA}
\affiliation{Princeton University, Princeton, New Jersey 08544, USA}
\affiliation{State University of New York, Buffalo, New York 14260, USA}
\affiliation{University of Rochester, Rochester, New York 14627, USA}
\affiliation{State University of New York, Stony Brook, New York 11794, USA}
\affiliation{Brookhaven National Laboratory, Upton, New York 11973, USA}
\affiliation{Langston University, Langston, Oklahoma 73050, USA}
\affiliation{University of Oklahoma, Norman, Oklahoma 73019, USA}
\affiliation{Oklahoma State University, Stillwater, Oklahoma 74078, USA}
\affiliation{Brown University, Providence, Rhode Island 02912, USA}
\affiliation{University of Texas, Arlington, Texas 76019, USA}
\affiliation{Southern Methodist University, Dallas, Texas 75275, USA}
\affiliation{Rice University, Houston, Texas 77005, USA}
\affiliation{University of Virginia, Charlottesville, Virginia 22904, USA}
\affiliation{University of Washington, Seattle, Washington 98195, USA}
\author{V.M.~Abazov} \affiliation{Joint Institute for Nuclear Research, Dubna, Russia}
\author{B.~Abbott} \affiliation{University of Oklahoma, Norman, Oklahoma 73019, USA}
\author{B.S.~Acharya} \affiliation{Tata Institute of Fundamental Research, Mumbai, India}
\author{M.~Adams} \affiliation{University of Illinois at Chicago, Chicago, Illinois 60607, USA}
\author{T.~Adams} \affiliation{Florida State University, Tallahassee, Florida 32306, USA}
\author{G.D.~Alexeev} \affiliation{Joint Institute for Nuclear Research, Dubna, Russia}
\author{G.~Alkhazov} \affiliation{Petersburg Nuclear Physics Institute, St. Petersburg, Russia}
\author{A.~Alton$^{a}$} \affiliation{University of Michigan, Ann Arbor, Michigan 48109, USA}
\author{A.~Askew} \affiliation{Florida State University, Tallahassee, Florida 32306, USA}
\author{S.~Atkins} \affiliation{Louisiana Tech University, Ruston, Louisiana 71272, USA}
% ...
%
%
\collaboration{The D0 Collaboration\footnote{with visitors from
$^{a}$Augustana College, Sioux Falls, SD, USA,
$^{b}$The University of Liverpool, Liverpool, UK,
$^{c}$UPIITA-IPN, Mexico City, Mexico,
$^{d}$DESY, Hamburg, Germany,
$^{e}$SLAC, Menlo Park, CA, USA,
$^{f}$University College London, London, UK,
$^{g}$Centro de Investigacion en Computacion - IPN, Mexico City, Mexico,
$^{h}$ECFM, Universidad Autonoma de Sinaloa, Culiac\'an, Mexico,
$^{i}$Universidade Estadual Paulista, S\~ao Paulo, Brazil,
$^{j}$Karlsruher Institut f\"ur Technologie (KIT) - Steinbuch Centre for Computing (SCC)
and
$^{k}$Office of Science, U.S. Department of Energy, Washington, D.C. 20585, USA.
}} \noaffiliation
\vskip 0.25cm

\date{4 March 2013}

\begin{abstract}
We perform a combination of searches for standard model Higgs boson production in \pp\ collisions
recorded by the D0 detector at the Fermilab Tevatron Collider
at a center of mass energy of $\sqrt{s}=1.96$~TeV.
The different production and decay channels have been analyzed separately, 
with integrated luminosities
of up to $\lumimax$~\ifb\ and for Higgs boson masses $90\leq M_H \leq 200$~GeV. 
We combine these final states to achieve optimal sensitivity
to the production of the Higgs boson. We also interpret the combination in terms of 
models with a fourth generation of fermions, and models with suppressed Higgs boson couplings to fermions.
The result excludes a standard model Higgs boson at 95\%\ C.L. 
in the ranges $90 < M_H < \excllow$~GeV 
and $\exclmin <M_H< \exclmax$~GeV, 
with an expected exclusion of
$\exclminexp <M_H<\exclmaxexp$~GeV.
In the range  $120 < M_H < 145$~GeV, the data exhibit an excess over the
expected background of up to two
standard deviations, consistent with the presence of a standard model Higgs boson of mass 125~GeV.

\end{abstract}

\pacs{14.80Bn, 13.85Rm}

\maketitle

\newpage
\section{Introduction}
\label{sec:intro}

\def\citeall{\cite{Abazov:2012wh97,Abazov:lvjets,Abazov:2012kg,Abazov:2012hv,Abazov:hWWdilep,Abazov:2012zj,Abazov:2013eha,Abazov:2012ee,Abazov:hgg}}

A fundamental goal of elementary particle physics is to understand the 
origin of electroweak symmetry breaking. 
The proposed mechanism in the standard model (SM) introduces 
a doublet of complex scalar fields into the SM Lagrangian, the neutral 
component of which develops a vacuum
expectation value that
generates the longitudinal polarizations and masses
 of the $W$ and $Z$ bosons. This mechanism~\cite{Higgs:1964ia,Englert:1964et,Higgs:1964pj,Guralnik:1964eu} 
gives rise to a single scalar boson, the Higgs boson ($H$), but does not provide a prediction for its
mass. Fermions acquire their masses via their interactions with the scalar field.
Precision electroweak data, including the latest $W$ boson 
and top quark mass
measurements at the CDF and D0
experiments at the Fermilab Tevatron 
Collider~\cite{Aaltonen:2012bp,Abazov:2012bv,Aaltonen:2012ra}, 
constrain the
mass of a SM Higgs boson to $M_H < 152$~GeV~\cite{bib:LEPEWWG} at 95\%\ confidence level 
(C.L.).  Direct searches at the ALEPH, DELPHI, L3, and OPAL experiments
at the CERN $e^+e^-$ Collider (LEP)~\cite{Barate:2003sz}, the CDF and D0 experiments~\cite{CDFandD0:2012aa,Aaltonen:2010sv}, and 
the ATLAS~\cite{Aad:2012an} and CMS~\cite{Chatrchyan:2012tx}
experiments at the CERN Large Hadron Collider (LHC) limit the SM Higgs boson 
mass to
122~GeV~$<M_H<127$~GeV at $95\%$~C.L. The ATLAS and CMS Collaborations have each observed a new boson 
in its bosonic decay modes with a mass near 125~GeV that is consistent with SM Higgs boson production~\cite{atlas-obs,cms-obs}. 
The CDF and D0 Collaborations have reported combined evidence
for a particle consistent with the SM Higgs boson produced in association with a $W$ or $Z$ boson that decays
to a $b\bar{b}$ pair~\cite{Aaltonen:2012qt}.

In this Article, we combine the results of direct searches for the SM Higgs
boson in \pp~collisions at~\tevE~recorded by the D0
experiment~\cite{Abachi:1993em,Abazov:2005pn,Abolins:2007yz,Angstadt:2009ie}.
The analyses combined here search for signals of
Higgs boson production through gluon-gluon fusion (GGF)
($gg\rightarrow H$), in association with vector bosons
(\pvh, where $V=W, Z$), and through virtual vector boson
fusion (VBF) (\vbf). The analyses
utilize data corresponding to integrated luminosities of up to
$\lumimax$~\ifb , collected during the years
2002--2011. The Higgs boson decay modes examined are
$H\rightarrow b{\bar{b}}$, $H\rightarrow W^+W^-$, 
$H\rightarrow \tau^+\tau^-$, and $H\rightarrow \gamma\gamma$.
We organize the searches into analysis subchannels comprising
different production, decay, and final state particle configurations,
designed to maximize the sensitivity for each particular Higgs boson
production and decay mode.  

We present an overview of the individual analyses
in Section~\ref{sec:channels}. Section~\ref{sec:bg} discusses the common methods 
of background estimation and simulation, while Section~\ref{sec:theory} details 
the signal predictions and associated uncertainties used in the analyses. In 
Section~\ref{sec:limits} we describe the statistical tehniques used in the combination,
and provide an overview of the most important systematic uncertainties. We 
validate our analysis techniques and statistical methods in Section~\ref{sec:diboson} by performing 
mesurements of the $WZ+ZZ$ and $WW$ production cross sections. In Section~\ref{sec:results} we present
our results for the SM Higgs boson as well as two interpretations beyond the SM. 
We summarize our results in Section~\ref{sec:conclusions}.

\section{Contributing Analyses}
\label{sec:channels}

\begin{table*}[htb]
\caption{\label{tab:chans}List of analysis channels, with the 
corresponding integrated luminosities and ranges in $M_H$ considered 
in the combined analysis. See Section~\ref{sec:channels} for details. We group the analyses in 
four categories, corresponding to the Higgs boson decay mode to which the analysis is most 
sensitive: $H\to b\bar{b}$, $H\to W^+W^-$, $H\to\tau^+\tau^-$, and $H\to\gamma\gamma$.}
\begin{ruledtabular}
\begin{tabular}{lcccc}
\\
Channel ($V=W,Z$ and $\ell=e, \mu$) & & Luminosity (\ifb)& $M_H$ (GeV) & Reference\\\hline
\whl         & & 9.7 & 90--150   & \cite{Abazov:2012wh97,Abazov:lvjets}\\
\zhl         & $H\to b\bar{b}$ & 9.7 & 90--150   & \cite{Abazov:2012kg,Abazov:2013mla}\\
\zhv         & & 9.5 & 100--150  & \cite{Abazov:2012hv}\\
\hline
\hwwlvlv     & & 9.7 & 100--200  & \cite{Abazov:hWWdilep}\\
\hwwmvtv     & & 7.3 & 155--200  & \cite{Abazov:2012zj}\\
\hwwlnuqq    &\multirow{2}{*}{$H\to W^+W^-$} & 9.7 & 100--200  & \cite{Abazov:lvjets}\\
\trilep      & & 9.7 & 100--200  & \cite{Abazov:2013eha}\\
\ssem        & & 9.7 & 100--200  & \cite{Abazov:2013eha}\\
\lnuqqqq     & & 9.7 & 100--200  & \cite{Abazov:lvjets}\\
\hline
\ttm         &\multirow{2}{*}{$H\to \tau^+\tau^-$} & 8.6 & 100--150  & \cite{Abazov:2013eha}\\
\tautaujj    & & 9.7 & 105--150  & \cite{Abazov:2012ee} \\
\hline
\hgg         & & 9.7 & 100--150  & \cite{Abazov:hgg} \\
\end{tabular}
\end{ruledtabular}
\end{table*}

A list of the analyses used in this combination is given in
Table~\ref{tab:chans}. We summarize the analyses below, grouping them according to the Higgs boson decay mode
to which the analysis is most sensitive. To facilitate their combination,
the analyses are constructed to be mutually exclusive
after all event selections.
\subsection{\bm{$H\rightarrow b\bar{b}$}  Analyses}
The most sensitive analyses for masses below
$M_H \lesssim 130$~GeV are those searching for \hbb\ decays in association with a
leptonically decaying $V$ boson.  To enhance the
\hbb\ component in the data, the analyses use an algorithm ($b$-tagger) to identify jets that are
consistent with $b$-quark lifetime and fragmentation.
Several kinematic variables sensitive to displaced vertices and to
tracks with large transverse impact parameters relative to the
production vertex are combined in a $b$-tagging discriminant.
This algorithm provides improvements when compared to
the previously used artificial neural network (ANN) $b$-tagger~\cite{Abazov:2010ab}.  
By adjusting the minimum requirement on the output of the $b$-tagger, a range of signal efficiencies and purities is achieved.

The D0 collaboration previously published a combination of $H\rightarrow b\bar{b}$ analyses on the full Run II dataset~\cite{Abazov:2012tf}. The two
searches focused on $ZH$ production described below are unchanged from the previous combination, while the $WH$ search
differs slightly from the previous iteration in the multijet background estimation and a refined treatment of some systematic uncertainties.

The \whl\ ($\ell=e,\mu$) analysis~\cite{Abazov:2012wh97,Abazov:lvjets} 
requires  topologies with a charged lepton, 
significant imbalance in the transverse energy (\met), and 
two or three jets ($j$).
A boosted decision tree (BDT) discriminant~\cite{narsky-0507157,Breiman1984,schapire01boostapproach,schapireFreund,friedman}
from {\sc tmva}~\cite{Hocker:2007ht}
 is used to discriminate against multijet
background.  Using the average of the two highest 
outputs from the $b$-tagger for
all selected jets, six mutually exclusive $b$-tagging categories are defined.  
Events with no $b$-tagged jets, and with exactly one of the lowest purity which can originate from a $c$ quark 
in the hadronic decay $W\to c\bar{s}$
are used for the
\hwwlnuqq\ analysis, while the
remaining events belong to 
the four $b$-tagging categories that are used in the \whl\ analysis.
A BDT discriminant is constructed 
for each lepton flavor, jet multiplicity, and $b$-tagging category.  
In addition
to kinematic variables, the inputs to the 
final discriminants include the $b$-tagger output and the output from the multijet discriminant.

The \zhl\ analysis~\cite{Abazov:2012kg,Abazov:2013mla} requires two isolated
charged leptons and at least two jets, at least one of which
must pass a tight $b$-tagging requirement.  A kinematic fit corrects the 
measured jet energies to their best fit values according to the constraints
that the dilepton invariant mass should be consistent with the $Z$ boson mass
$M_Z$ and
the total transverse momentum of the leptons and jets 
should be consistent with zero.
The events are divided
into ``double-tag'' and ``single-tag'' subchannels depending on
whether a second jet passes a loose $b$-tagging requirement.
The analysis uses random forest (RF)~\cite{Hocker:2007ht} discriminants to provide distributions
for the final statistical analysis, applied in a
two-step process.  First, the events are divided into independent
\ttbar-depleted and \ttbar-enriched subchannels using a dedicated RF that is
trained to discriminate signal from the \ttbar\ backgrounds in each
lepton and $b$-tagging subchannel.  Final discriminants are then
constructed to separate signal from all backgrounds.  The limit is
calculated using the output distributions of the final discriminants
for both the \ttbar-depleted and \ttbar-enriched samples.
The \tautaujj\ analysis, where $\tau_{h}$ denotes 
$\tau$-lepton decays into hadrons,
 discussed in Sec.~\ref{sec:taugamma} includes a contribution from $ZH$ 
production with $Z\to\tau^+\tau^-$ and $H\to b\bar{b}$ decays.

The \zhv\ analysis~\cite{Abazov:2012hv} selects events with large \met\ and
two jets. 
This search is also
sensitive to the $WH$ process when the charged lepton
from $W\to\ell\nu$ decay is not identified.
Events selected in
the \whl\ analysis are rejected to ensure no overlap
between the two analyses. About 47\% of signal in this 
analysis comes from \whl\ events in which the charged lepton fails
the \whl\ analysis selection requirements. 
Variables such as \met\ significance  
and a track-based missing transverse momentum are used to reject events 
with \met\ arising from mismeasurement of jet energies.
The multijet background is further
reduced by employing a dedicated BDT discriminant before applying
$b$-tagging.  
Two $b$-tagging subchannels are defined using the sum of the 
$b$-tagging discriminant outputs of the two jets.
 BDT classifiers, trained separately for 
different $b$-tagging categories, are used as a final discriminant.

\subsection{\bm{$H\rightarrow WW^*$} Analyses}
We search for Higgs boson decays to two $W$ bosons from the three
dominant production mechanisms: gluon-gluon fusion, associated
production, and vector boson fusion.  In \hww\ decays with $M_H<2 M_W$,
at least one of the $W$ bosons will be virtual ($W^*$).  

The dominant search channels are \hwwlnulnu~\cite{Abazov:hWWdilep}.  
The presence
of neutrinos in the final state prevents precise reconstruction
of the candidate $M_H$.  
Events are characterized by
large \met\ and two isolated leptons of opposite electric charge.  
Each final state is further
subdivided according to the number of jets in the event: no jets, one, and
more than one jet.  
This division requires an evaluation of
theoretical uncertainties on the signal predictions for each jet 
category, as will be discussed in Section~\ref{sec:theory}.

The dielectron and dimuon channels use BDT discriminants to
reduce the dominant Drell-Yan background, while the $e^\pm \mu^\mp$
channel uses $\met$-related variables to minimize backgrounds.  
All channels separate events into $WW$-enriched
and $WW$-depleted subchannels. In the dielectron and dimuon channels, dedicated
BDTs are applied to events with no jets or exactly one jet. Events with no jets are 
split according to the lepton quality in the $e^\pm \mu^\mp$ channel.
BDT response distributions, using several kinematic
variables as inputs, are used as final discriminants.
Inputs also include $b$-tagging information for
subchannels containing jets to reject the \ttbar\ background.

We consider final states where at least one $W$ boson decays to $\tau\nu$, and the $\tau$ 
lepton decays into hadrons ($\tau_{h}$) and $\nu_{\tau}$
(\hwwmvtv)~\cite{Abazov:2012zj}.
Final states involving other $\tau$ decays and misidentified $\tau_h$
decays are included in the \hww\
analyses channels.  This channel uses ANN outputs~\cite{Hocker:2007ht} 
for a final discriminant. 

We also include analyses that search for \hww\ with one of the $W$
bosons decaying into $\bar{q}q^{\prime}$.  The \hwwlnuqq\ analysis~\cite{Abazov:lvjets}
has the same initial selections as the \whl\ search, 
except that it considers only
events with no $b$-tagged jets, and with exactly one $b$-tagged jet of the lowest purity that can originate from a $c$ quark.
The RF discriminants trained for each lepton flavor, jet multiplicity, 
and $b$-tagging category serve as the final discriminant variables.

For \vhvww\ production, we consider final states containing: (i) three charged
leptons (\trilep)~\cite{Abazov:2013eha};
(ii) an electron and muon with the same charge
($e^\pm\mu^\pm+X$)~\cite{Abazov:2013eha};  and (iii) final states with one lepton, \met\ and at least four jets (\lnuqqqq)~\cite{Abazov:lvjets}.

The \trilep\ analyses use
BDT outputs as final discriminants. In the $\mu\mu e$
final state, events are split into three mutually exclusive regions to separate signal from
$Z$+jets and other backgrounds.

The $e^\pm\mu^\pm+X$ analysis, in which the same-sign requirement
suppresses the Drell-Yan background, uses a two-step multivariate approach: 
(i) a BDT is used to suppress
most of the dominant backgrounds from multijet,
$W+\text{jets}$, and $W+\gamma$
events, and (ii) another BDT is used to discriminate signal
from the remaining backgrounds.

The \lnuqqqq\ analysis~\cite{Abazov:lvjets} has selections
similar to the 
\hwwlnuqq\ analysis, but requires at least four jets. 
Separate BDTs are trained for different backgrounds, and then they are used as input variables to the final RF discriminant.

\subsection{\bm{\htt} and \bm{\hgg} analyses\label{sec:taugamma}}

Higgs boson decays involving $\tau$ leptons are included in different ways. 
The \ttm\ analysis~\cite{Abazov:2013eha} 
uses a two-stage BDT approach, in which the
first BDT discriminates between signal and
backgrounds other than diboson ($VV$) production, and the second
BDT, trained to distinguish between signal
and all backgrounds, is implemented 
after selecting events that pass the first BDT
requirement.

The \tautaujj\ analysis~\cite{Abazov:2012ee} selects 
events with one electron or muon, a $\tau_h$, and 
two or more jets. It is sensitive to associated $VH$, VBF, 
and $gg\to H+X$ production, and to both $H\rightarrow \tau\tau$ 
and $H\rightarrow WW$ decays. A BDT, trained to 
distinguish between signal with $H\to\tau\tau$ and $H \to WW$ 
decays, is used to create  $\tau\tau$- and $WW$-dominated subchannels
within the electron and muon channels.
Each of the four resulting subchannels
has a BDT
as the final discriminant. 

We also include in the combination
an analysis that searches for Higgs boson
decaying to two photons~\cite{Abazov:hgg}. The Higgs boson is
assumed to be
produced via GGF, VBF, 
and associated $VH$ production.
The contribution of jets misidentified as photons is reduced by
combining information sensitive to differences in the energy
deposition in the tracker, calorimeter and
central preshower in an ANN for each photon candidate.  The ANN output
defines photon-dominated and jet-dominated regions, 
each of which is split into
signal-rich and signal-depleted contributions based on
 the diphoton invariant mass. 
A BDT built with ten variables, including the diphoton mass, serves as the final discriminant in the signal rich region, while the diphoton mass only is the final discriminant in signal-depleted regions.

\section{Background Estimation\label{sec:bg}}

All analyses estimate backgrounds from multijet production
through special data control samples. 
The other backgrounds are determined from Monte Carlo (MC) simulation. MC 
samples are generated using the {\sc pythia}~\cite{Sjostrand:2006za}, 
{\sc alpgen}~\cite{Mangano:2002ea}, 
{\sc sherpa}~\cite{Gleisberg:2008ta}, or 
{\sc singletop}~\cite{Boos:2004kh,Boos:2006af} event generators,
with {\sc pythia} also providing parton showering and hadronization for
{\sc alpgen} and {\sc singletop}. 
All generators use the {\sc CTEQ6L1}~\cite{Lai:1996mg,Nadolsky:2008zw} leading
order (LO) parton distribution functions (PDF). Drell-Yan and $W$+jets yields are normalized to
next-to-next-to-LO (NNLO) calculations~\cite{Hamberg:1990np}, or, in some analyses, to 
data control samples~\cite{Abazov:2012kg,Abazov:2013mla,Abazov:hWWdilep,Abazov:2013eha}.
For the $V+b\bar{b}/c\bar{c}$ MC samples, generated separately
from the $V$+light-flavor events,
we apply additional normalization factors calculated at next-to-LO (NLO) from {\sc mcfm}~\cite{Campbell:1999ah,mcfm_code} to account 
for the heavy-flavor to light-flavor production ratio. 
Diboson background cross sections are normalized to NLO calculations from 
{\sc mcfm}. Top quark pair and single top quark production are normalized to approximate
NNLO~\cite{Langenfeld:2009wd} and
next-to-NNLO (NNNLO)~\cite{Kidonakis:2006bu} calculations, respectively. 
We correct
the transverse momentum ($p_T$) spectrum of the $Z$ boson in the MC
to match that observed in data~\cite{Abazov:2007nt}. We correct
the $W$ boson $p_T$ using the same dependence,
taking into account differences between the $p_T$ spectra of
$Z$ and $W$ bosons predicted in NNLO QCD~\cite{Melnikov:2006kv}.
We account for $W\gamma^*$ production and its interference
with $WZ$ production 
using {\sc powheg}~\cite{powheg} in analyses where this effect is significant: \hwweemm, \ssem, and \trilep.

\section{Signal Predictions and Uncertainties\label{sec:theory}}

An outline of the procedures for the signal predictions and associated uncertainties is given below.
Reference~\cite{CDFandD0:2012aa} contains
a more complete discussion.  

We simulate signal with {\sc pythia} using the {\sc CTEQ6L1}
PDFs to model the parton shower, fragmentation, and hadronization.  
We reweight the Higgs boson $p_T$ spectra for GGF production
to the prediction obtained from 
{\sc hqt}~\cite{Bozzi:2003jy,Bozzi:2005wk,deFlorian:2011xf}.
To evaluate the
impact of the scale uncertainty on the differential spectra, we use
the {\sc resbos}~\cite{Balazs:2000sz,Cao:2009md} generator and apply the
scale-dependent differences in the Higgs boson $p_T$ spectrum to the
{\sc hqt} prediction. We propagate these changes to the final discriminants
as a systematic uncertainty on the differential distribution
which is included in the calculation of the limits.

We normalize the Higgs boson signal predictions to the most recent
higher-order calculations (see Table~\ref{tab:higgsxsec}).  
The $gg\rightarrow H$ production
cross section (\sigmaggh) is calculated at NNLO
in QCD with a next-to-next-to-leading-log resummation of
soft gluons. The calculation also includes two-loop electroweak
effects and the running $b$-quark
mass~\cite{Anastasiou:2008tj,deFlorian:2009hc}. The values in
Table~\ref{tab:higgsxsec} are updates~\cite{grazziniprivate} of these
predictions, with the top quark mass set to 173.1~GeV~\cite{Group:2009ad}, and includes an
exact treatment of the massive top quark and bottom quark loop corrections up to
NLO and next-to-leading-log (NLL) accuracy.  The factorization scale $\mu_F$ and
renormalization scale $\mu_R$ choices for this calculation are
$\mu_F=\mu_R=M_H$. These calculations are improvements over the previous
NNLO calculations of \sigmaggh~\cite{Harlander:2002wh,Anastasiou:2002yz,Ravindran:2003um}.
We apply the electroweak corrections computed in
Refs.~\cite{Actis200812,Aglietti:2006yd}. The soft gluon resummation uses the
calculations of Ref.~\cite{Catani:2003zt}.  The gluon PDF and the
accompanying value of $\alpha_s(q^2)$ strongly influence \sigmaggh.  The cross sections we use
are calculated with the MSTW~2008 NNLO PDFs~\cite{Martin:2009bu}, as
recommended by the PDF4LHC working group~\cite{Alekhin:2011sk,Botje:2011sn}.  

\begin{table*}
\begin{center}
\caption{
The production cross sections (in fb) and decay branching fractions
(in \%) for each SM Higgs boson mass considered in the combined analysis.
}
\vspace{0.2cm}
\label{tab:higgsxsec}
\begin{ruledtabular}
{\footnotesize
\begin{tabular}{crrrrcccccc}
$M_H$~(GeV)& $\sigma_{gg\rightarrow H}$  & $\sigma_{WH}$  & $\sigma_{ZH}$  & $\sigma_{VBF}$  & $\mathcal{B}(H\rightarrow b{\bar{b}})$ & $\mathcal{B}(H\rightarrow c{\bar{c}})$ & $\mathcal{B}(H\rightarrow \tau^+{\tau^-})$ & $\mathcal{B}(H\rightarrow W^+W^-)$ & $\mathcal{B}(H\rightarrow ZZ)$ & $\mathcal{B}(H\rightarrow \gamma \gamma)$\\ 
\hline
   90  &   2442   &   394.7  &   224.0   &    118.2 &  81.2   & 3.78     & 8.41    & 0.21   & 0.042  & 0.123    \\ 
   95  &   2101   &   332.1  &   190.3   &    108.8 &  80.4   & 3.73     & 8.41    & 0.47   & 0.067  & 0.140    \\
   100 &   1821   &   281.1 &   162.7   &    100.2 &  79.1   & 3.68     & 8.36    & 1.11   & 0.113  & 0.159    \\ 
   105 &   1584   &   238.7 &   139.5   &     92.3 &  77.3   & 3.59     & 8.25    & 2.43   & 0.215  & 0.178    \\
   110 &   1385   &   203.7 &   120.2   &     85.2 &  74.5   & 3.46     & 8.03    & 4.82   & 0.439  & 0.197    \\
   115 &   1215   &   174.5 &   103.9   &     78.7 &  70.5   & 3.27     & 7.65    & 8.67   & 0.873  & 0.213    \\
   120 &   1072   &   150.1 &    90.2   &     72.7 &  64.9   & 3.01     & 7.11    & 14.3   & 1.60   & 0.225    \\
   125 &    949   &   129.5 &    78.5   &     67.1 &  57.8   & 2.68     & 6.37    & 21.6   & 2.67   & 0.230    \\
   130 &    842   &   112.0 &    68.5   &     62.1 &  49.4   & 2.29     & 5.49    & 30.5   & 4.02   & 0.226    \\
   135 &    750   &    97.2 &    60.0   &     57.5 &  40.4   & 1.87     & 4.52    & 40.3   & 5.51   & 0.214    \\
   140 &    670   &    84.6 &    52.7   &     53.2 &  31.4   & 1.46     & 3.54    & 50.4   & 6.92   & 0.194    \\
   145 &    600   &    73.7 &    46.3   &     49.4 &  23.1   & 1.07     & 2.62    & 60.3   & 7.96   & 0.168    \\
   150 &    539   &    64.4 &    40.8   &     45.8 &  15.7   & 0.725    & 1.79    & 69.9   & 8.28   & 0.137    \\
   155 &    484   &    56.2 &    35.9   &     42.4 &  9.18   & 0.425    & 1.06    & 79.6   & 7.36   & 0.100    \\
   160 &    432   &    48.5 &    31.4   &     39.4 &  3.44   & 0.159    & 0.397   & 90.9   & 4.16   & 0.0533   \\
   165 &    383   &    43.6 &    28.4   &     36.6 &  1.19   & 0.0549   & 0.138   & 96.0   & 2.22   & 0.0230   \\
   170 &    344   &    38.5 &    25.3   &     34.0 &  0.787  & 0.0364   & 0.0920  & 96.5   & 2.36   & 0.0158   \\
   175 &    309   &    34.0 &    22.5   &     31.6 &  0.612  & 0.0283   & 0.0719  & 95.8   & 3.23   & 0.0123   \\
   180 &    279   &    30.1 &    20.0   &     29.4 &  0.497  & 0.0230   & 0.0587  & 93.2   & 6.02   & 0.0102   \\
   185 &    252   &    26.9 &    17.9   &     27.3 &  0.385  & 0.0178   & 0.0457  & 84.4   & 15.0   & 0.00809  \\
   190 &    228   &    24.0 &    16.1   &     25.4 &  0.315  & 0.0146   & 0.0376  & 78.6   & 20.9   & 0.00674  \\
   195 &    207   &    21.4 &    14.4   &     23.7 &  0.270  & 0.0125   & 0.0324  & 75.7   & 23.9   & 0.00589  \\
   200 &    189   &    19.1 &    13.0   &     22.0 &  0.238  & 0.0110   & 0.0287  & 74.1   & 25.6   & 0.00526  \\ 
\end{tabular}	
}
\end{ruledtabular}	
\end{center}	
\end{table*}

For analyses that consider inclusive $gg\rightarrow H$ production, but
do not split the signal into separate channels based on the
number of reconstructed jets, we use the uncertainties on inclusive production from
the simultaneous variation of the factorization and renormalization
scale up and down by a factor of two.  We use the prescription of the
PDF4LHC working group for evaluating PDF uncertainties on the
inclusive production cross section.  QCD scale uncertainties that
affect the cross section via their impact on the PDFs are included as
a correlated part of the total scale uncertainty.  The remainder of
the PDF uncertainty is treated as uncorrelated with the uncertainty on 
the QCD scale.

For analyses of $gg\rightarrow H$ production that divide events
into separate channels based on the number of reconstructed jets, we evaluate 
the impact of the scale uncertainties
following the procedure of Ref.~\cite{Stewart:2011cf}.  We treat as uncorrelated the QCD scale uncertainties
obtained from the NNLL inclusive~\cite{deFlorian:2009hc,Anastasiou:2008tj},
NLO with one or more jets~\cite{Anastasiou:2009bt}, and NLO with two or
more jets~\cite{Campbell:2010cz} cross section calculations.  We then obtain QCD scale uncertainties
for the exclusive $gg\rightarrow H+n$~jets ($n=0,1,\geq 2$) categories 
by propagating the uncertainties on the inclusive cross
section predictions through the subtractions needed for the
exclusive rates.  For example, we obtain the $H+0$~jet cross section
by subtracting the NLO $H+ \geq 1$~jets cross section from the
inclusive NNLL+NNLO cross section. We therefore assign three
separate, uncorrelated QCD scale uncertainties that lead to correlated
and anticorrelated contributions between exclusive jet
categories. The procedure in Ref.~\cite{Anastasiou:2009bt} is used to
determine the uncertainties from the choice of PDF.  These are obtained separately for
each jet bin and treated as fully correlated between jet bins.

Another source of uncertainty in the prediction of
\sigmaggh\ is the extrapolation of QCD corrections
computed for heavy top-quark loops to the light-quark loops
included as part of the electroweak corrections.  Uncertainties at the
level of 1--2\% are already included in the cross section values we
use~\cite{deFlorian:2009hc,Anastasiou:2008tj}.  
The factorization of QCD corrections
is expected to be reliable for $M_H$ values much larger than the masses of the particles contributing to the loop~\cite{Anastasiou:2008tj}.
There is a 4\%\ change in the predicted cross
section when removing all QCD corrections from the diagrams
containing light-flavored quark loops.
For
the $b$-quark loop~\cite{Anastasiou:2008tj}, the QCD corrections are much
smaller than for top-quark loops, confirming that the procedure does
not introduce significant uncertainties. We therefore do not consider 
any additional uncertainties from this source.

For $WH$ and $ZH$ production we use cross
sections computed at NNLO~\cite{Baglio:2010um}.  This calculation
starts with the NLO calculation of {\sc v2hv}~\cite{spira_prog} and includes
NNLO QCD contributions~\cite{Brein:2003wg}, as well as one-loop
electroweak corrections~\cite{Ciccolini:2003jy}.  For VBF production,
we use the VBF cross section
computed at NNLO in QCD~\cite{Bolzoni:2011cu}.  Electroweak corrections to
the VBF production cross section, computed with the {\sc hawk}
program~\cite{Ciccolini:2003jy} are included although they
are very small ($\leq 0.03$~fb) for the
$M_H$ range that we consider.

The predictions of Higgs boson decay branching fractions, ${\cal B}$,
are taken from
{\sc hdecay}~\cite{Djouadi:1997yw, Butterworth:2010ym}, and are also listed in
Table~\ref{tab:higgsxsec}.
Uncertainties on the
branching fractions are taken from
Ref.~\cite{Baglio:2010ae}.  

\section{Limit Calculations \label{sec:limits}}

We combine results using the $CL_s$ method with a negative
log-likelihood ratio (LLR) test statistic~\cite{Junk:1999kv,Read:2002hq} for
the  signal-plus-background ($s+b$) and background-only ($b$) hypotheses, where
${\rm LLR} = -2\ln(L_{s+b}/L_{b})$, and $L_{hy}$ is the likelihood function for the 
hypothesis $hy$. The value of $CL_s$ is defined as $CL_s = CL_{s+b}/CL_b$ 
where $CL_{s+b}$ and $CL_b$ are the confidence levels for the $s+b$
and the $b$ hypotheses, respectively.  These
confidence levels are evaluated by integrating the corresponding LLR
distributions populated by simulating outcomes assuming Poisson
statistics. Separate channels and bins are combined by summing LLR
values over all bins and channels. This method provides a robust means
of combining channels while maintaining each individual channel's
sensitivity and different systematic uncertainties. Systematic
uncertainties are treated as nuisance parameters with Gaussian 
probability distributions
constrained by their priors. This approach ensures that 
the uncertainties and their correlations are propagated to the outcome 
with their appropriate weights. 

To minimize the degrading effects of systematic uncertainties on the
search sensitivity, we fit the individual background contributions
to the observed data by maximizing a likelihood function~\cite{wade_tm}. 
The likelihood is a joint Poisson probability
over the number of bins in the calculation and is a function of the
nuisance parameters and their uncertainties.
The maximization of the likelihood function
is performed over the nuisance parameters, with separate
fits performed to both
the $b$ and $s+b$ hypotheses for each Poisson MC trial. We have verified that
all fit parameters and pulls on the systematic uncertainties are well-behaved.

The $CL_s$ approach used in this combination utilizes binned
final variable distributions rather than a single fully
integrated value for each contributing analysis. The signal exclusion
criteria are determined by increasing the signal cross section until
$CL_s < 0.05$, which defines a signal cross section excluded at the
95\%\ C.L.

\subsection{Final Variable Distributions}

All analyses are performed for the $M_H$ range listed in 
Table~\ref{tab:chans} at 5~GeV intervals.  
Each analysis provides binned distributions of its 
final discriminants for each value of $M_H$ and subchannel.
The input distributions for individual channels 
can be found in the corresponding references in Table~\ref{tab:chans}.

The limit calculation uses the full information available in
 the individual discriminants. However, for visualization purposes
it can be useful to collect all of the inputs
into a single distribution.  To preserve sensitivity from the bins
with high signal-to-background ($s/b$) ratios, where $s$ is the number of signal and $b$ the number of background events, only bins with similar
$s/b$ ratio are combined.  The aggregate distribution is formed by
reordering all of the bins from the input distributions according to
$s/b$ ratio.  The range of $s/b$ ratio is large, so $\log_{10}(s/b)$ is
used.  Figure~\ref{fig:sbInputs} shows the aggregate distributions for
$M_H=125$~GeV and $M_H=165$~GeV, indicating
good agreement between data and predictions over several orders of
magnitude.  Figure~\ref{fig:subtractInputs} shows the same
distributions after subtracting the expected background from the data, where solid lines
represent the $\pm 1$
 standard deviations (s.d.) in systematic uncertainty 
after a fit to the background-only hypothesis. 
Integrating the distributions in Fig.~\ref{fig:sbInputs} from the highest to the lowest $s/b$ events
illustrates how the data compare to the $b$ and
$s+b$ hypotheses as the events in the highest $s/b$ bins
accumulate.  Figure~\ref{fig:integralInputs} shows these cumulative
distributions for approximately 150 of the most significant events as a
function of the integrated number of signal events.  
For $M_H=125$~GeV, the highest $s/b$ bins contain an excess
of signal-like events, while for $M_H=165$~GeV, the data
follow the background-only expectation.

\begin{figure*}[htbp]
\begin{centering}
\includegraphics[width=0.45\textwidth]{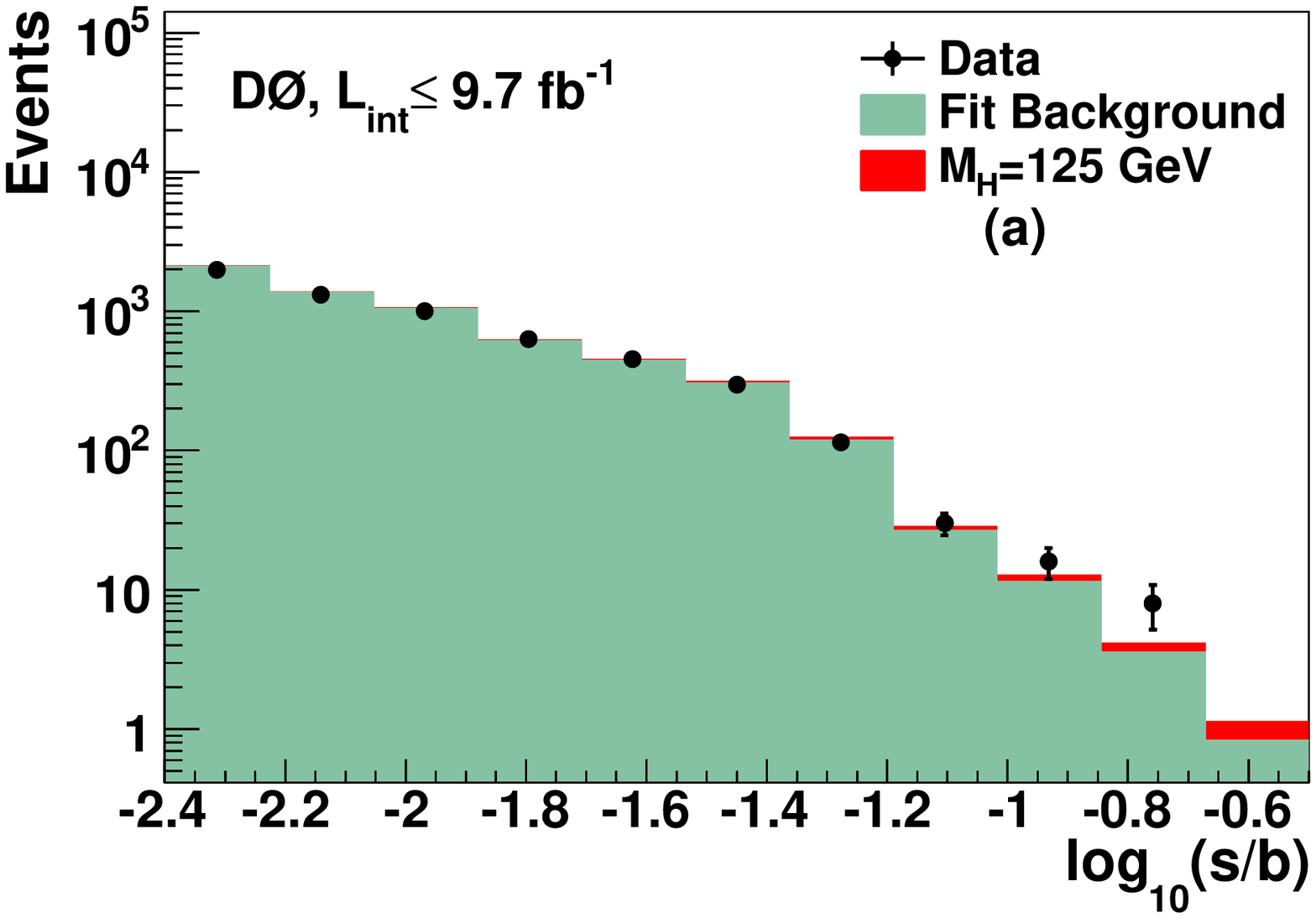}
\includegraphics[width=0.45\textwidth]{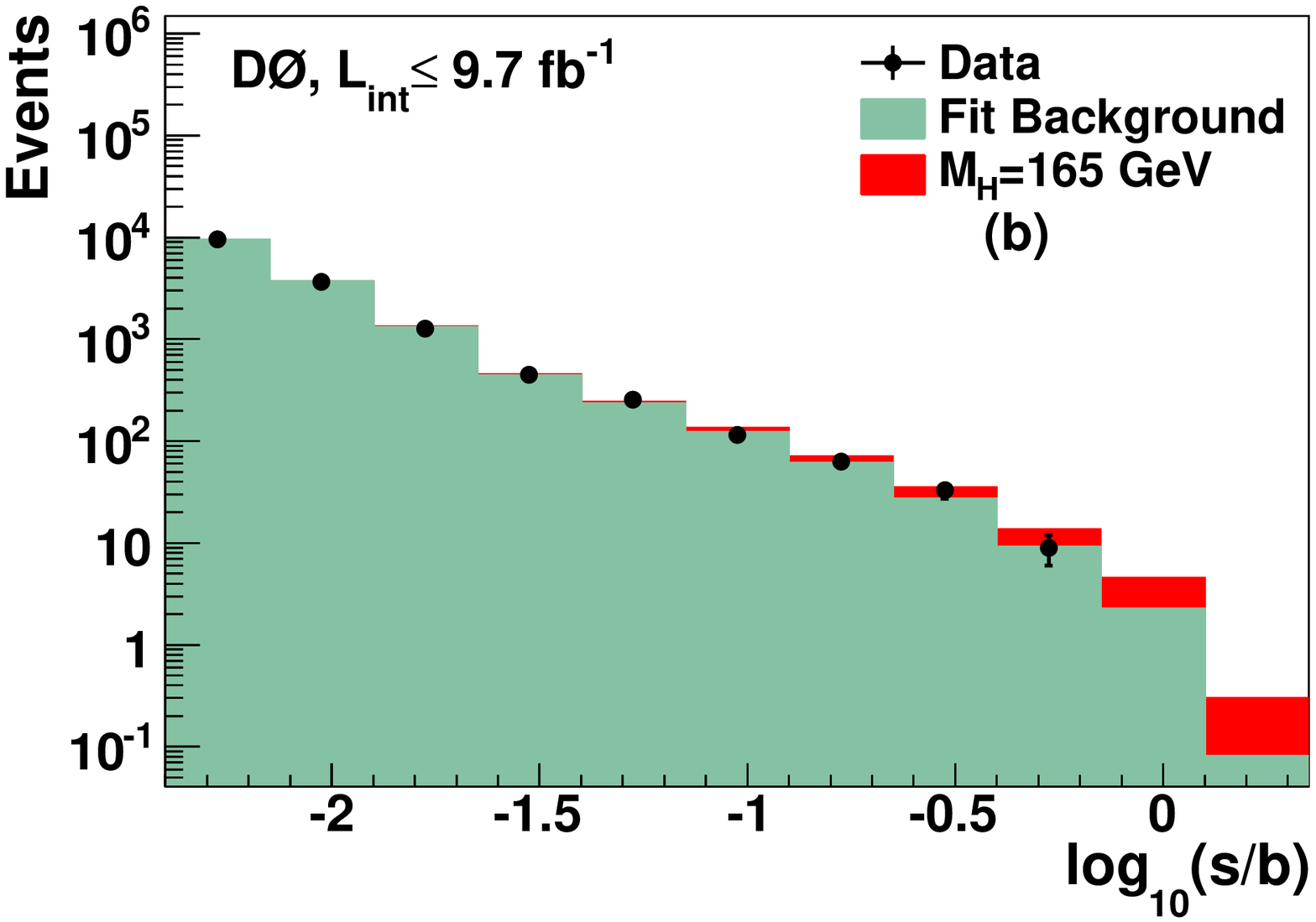}
\caption{
\label{fig:sbInputs} (color online) Distributions of $\log_{10}(s/b)$ for data from all
contributing channels for (a) $M_H=125$~GeV, and
(b) $M_H=165$~GeV after a fit to data assuming the background-only 
hypothesis. The data (points with Poisson statistical errors on the expected number of signal+background events) are compared to the expectation from background (light shaded) and signal (dark shaded). }
\end{centering}
\end{figure*}

\begin{figure*}[htbp]
\begin{centering}
\includegraphics[width=0.45\textwidth]{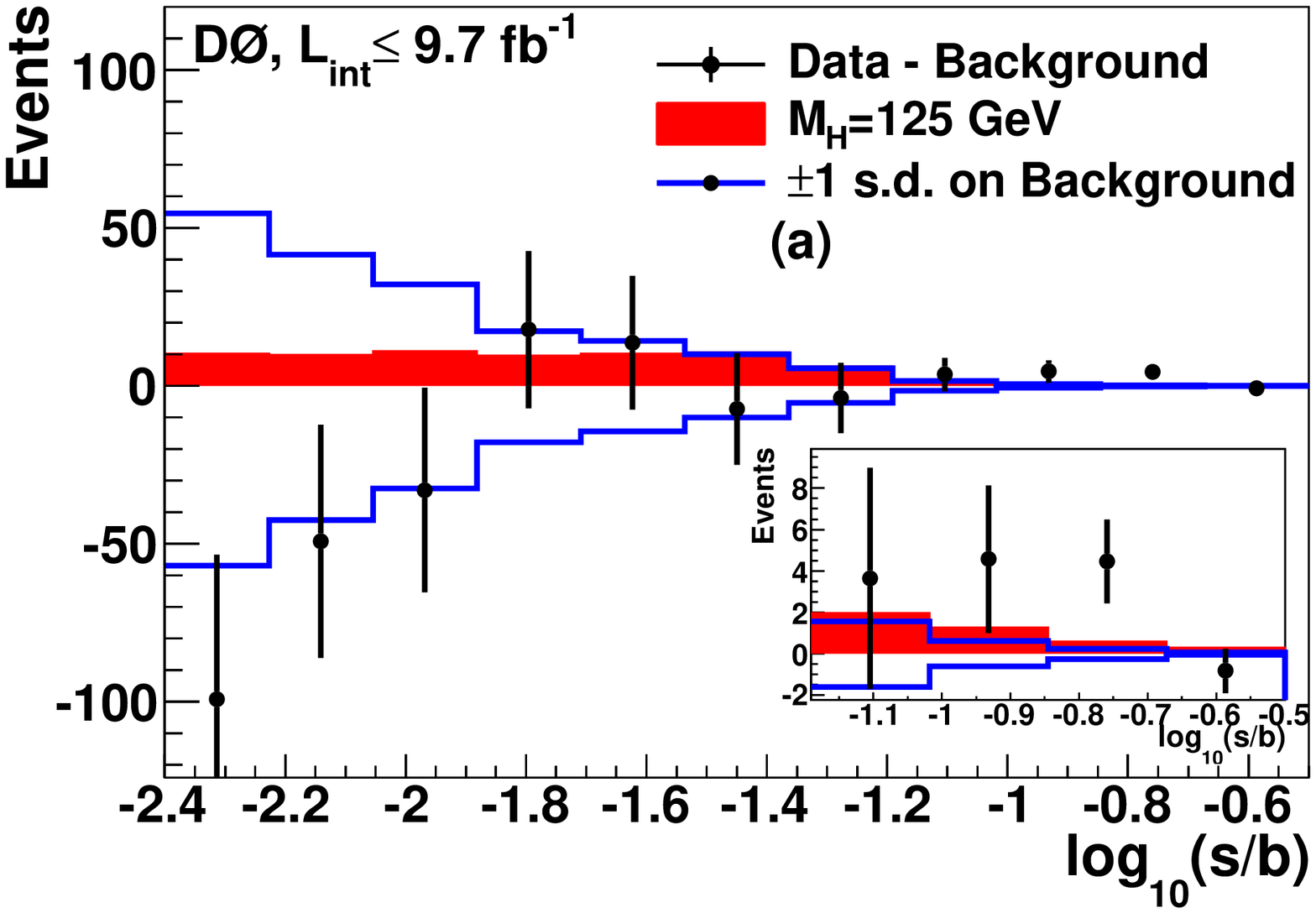}
\includegraphics[width=0.45\textwidth]{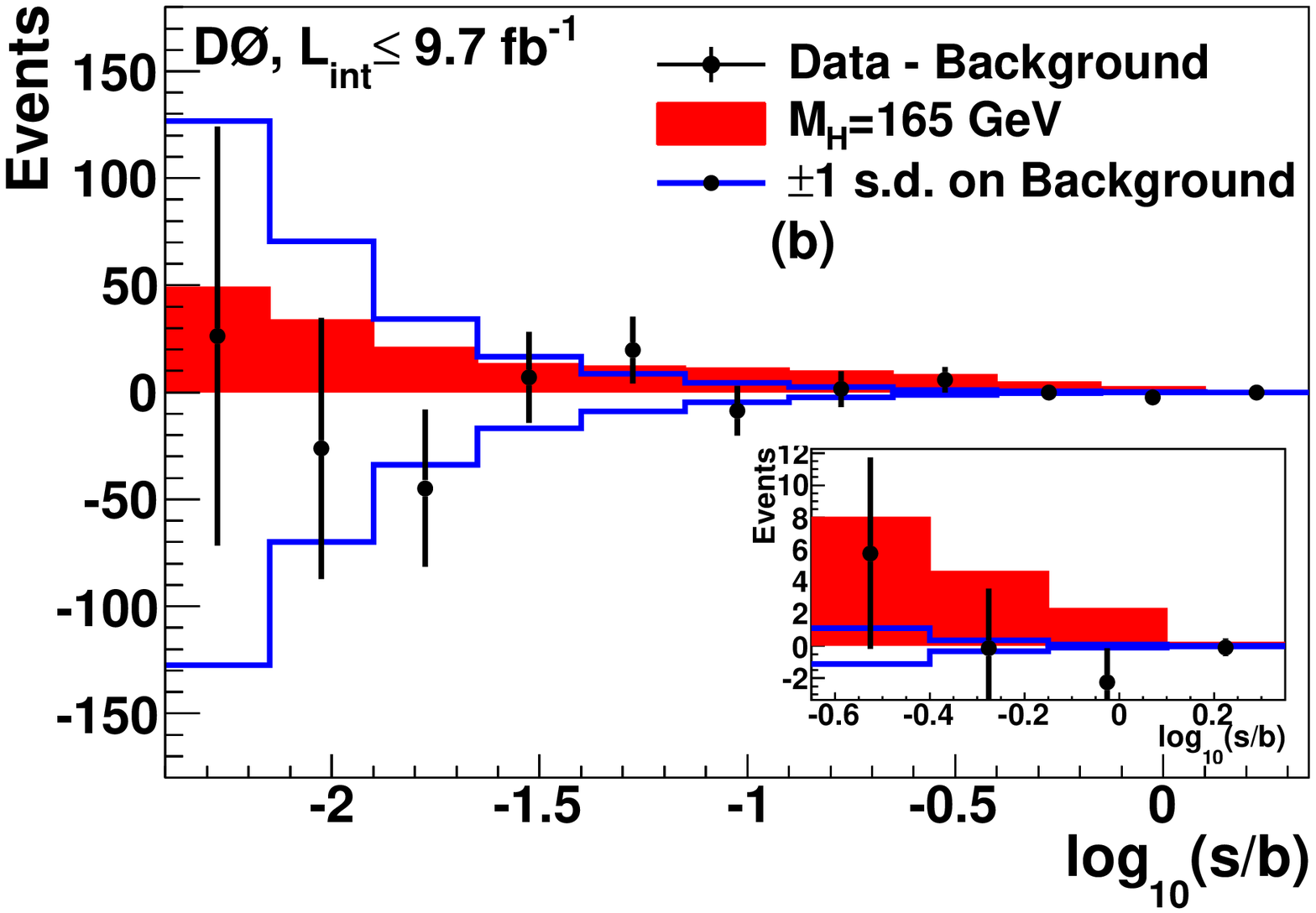}
\caption{
\label{fig:subtractInputs} (color online) Background-subtracted distributions as a function of $\log_{10}(s/b)$ 
for data from all contributing channels for (a) $M_H=125$~GeV and
(b) $M_H=165$~GeV after a fit to data assuming the background-only hypothesis.  The background-subtracted data (points with Poisson statistical errors on the expected number of signal+background events) 
are compared to the expected signal 
(shaded). The solid 
lines represent the $\pm 1$~s.d.~systematic uncertainty on the background after the fit.}
\end{centering}
\end{figure*}

\begin{figure*}[htbp]
\begin{centering}
\includegraphics[width=0.45\textwidth]{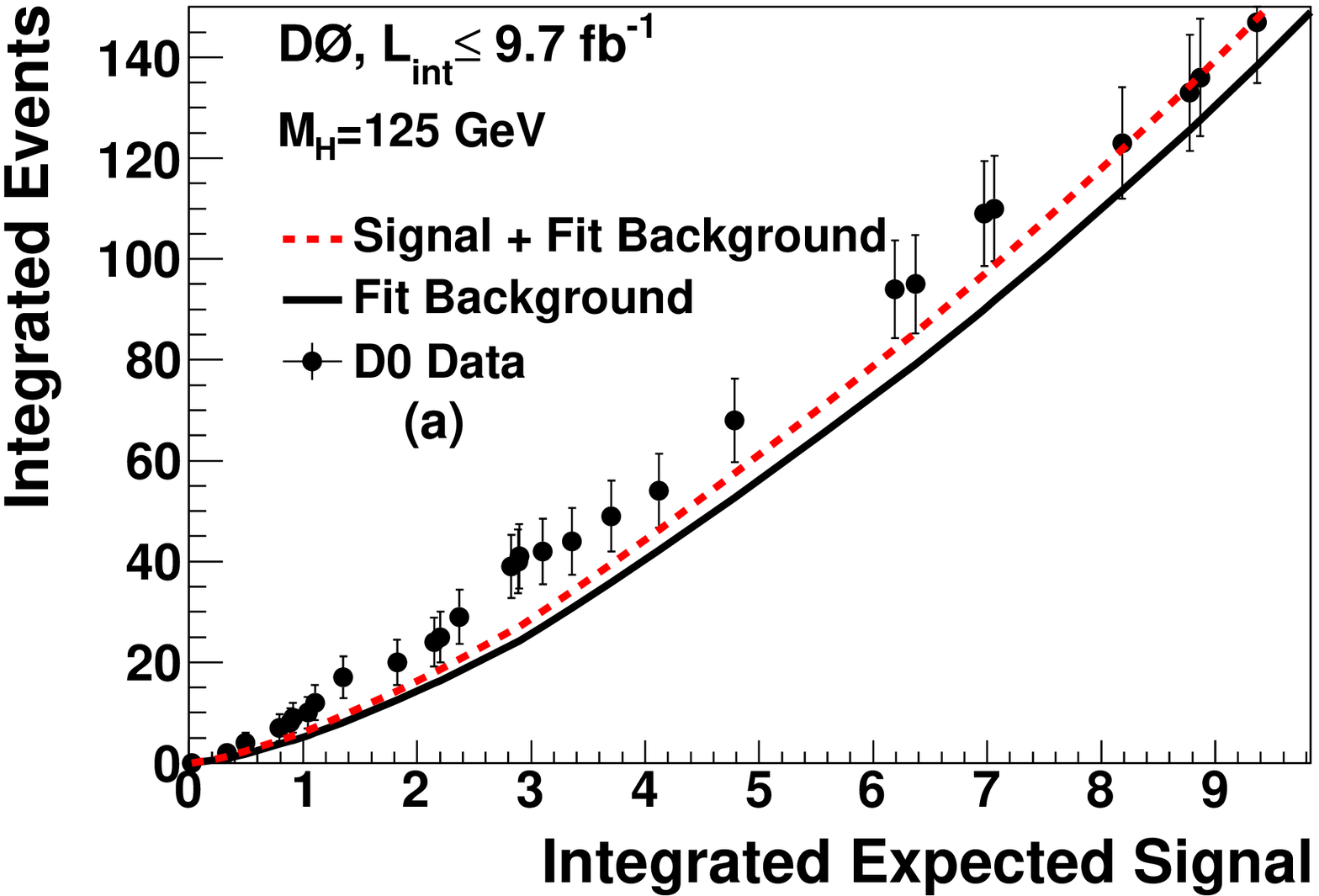}
\includegraphics[width=0.45\textwidth]{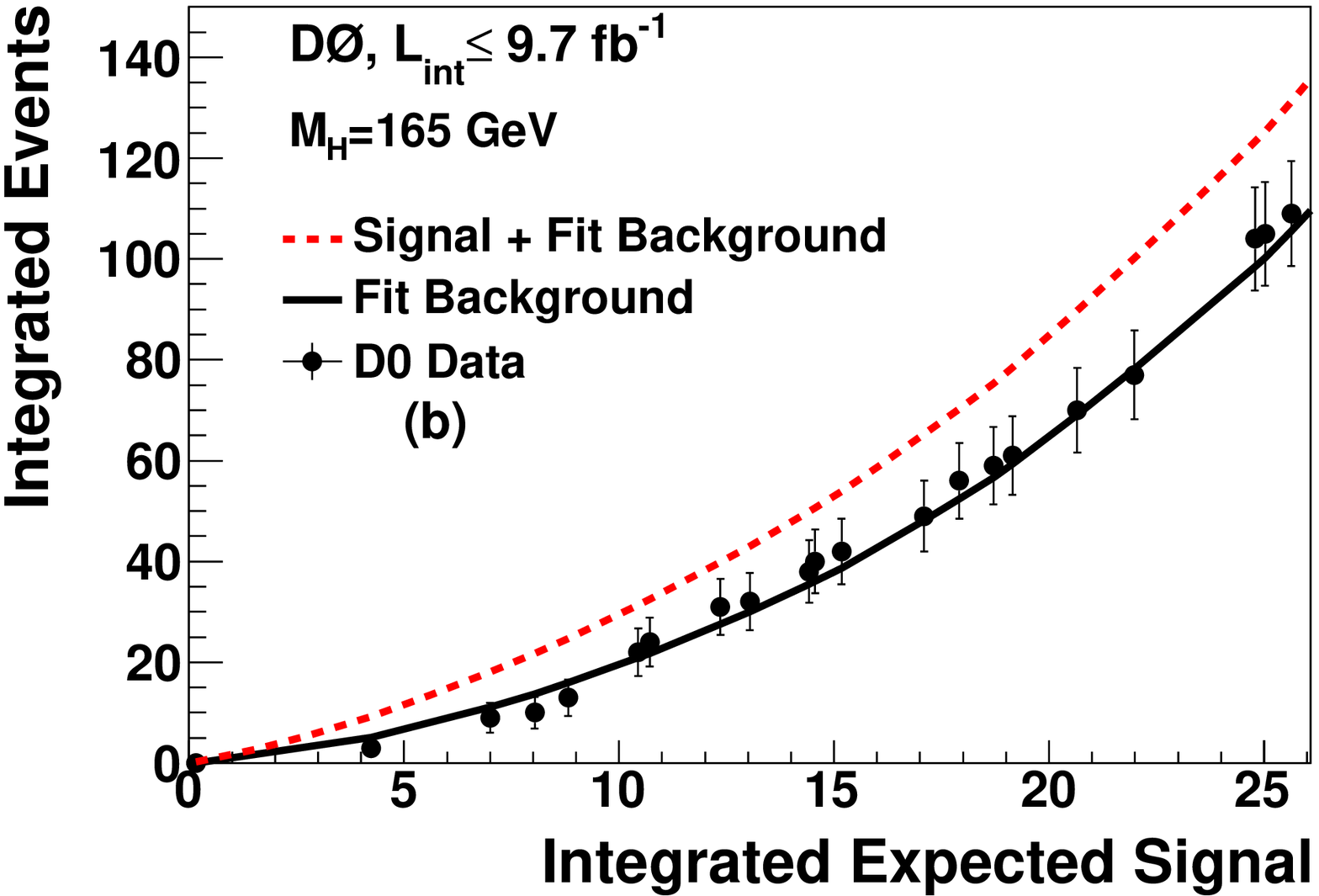}
\caption{
\label{fig:integralInputs} (color online) Cumulative number of events 
after integrating the final discriminant bins in decreasing order of $\log_{10} (s/b)$
until the expected signal yield indicated on the $x$-axis is reached for (a) $M_H=125$~GeV and
(b) $M_H=165$~GeV.  The integrated $b$-only and $s+b$ 
predictions are shown after their respective fits as a function of the accumulated number of
signal events.  The points show the integrated number of observed
events with statistical errors.
}
\end{centering}
\end{figure*}


\subsection{Systematic Uncertainties \label{sec:systs}}

Systematic uncertainties on signal and backgrounds
vary among the analyses
and they are described in detail in Refs.~\citeall.
We summarize below only the major components.
Most analyses have an uncertainty of 6.1\%\ from the integrated
luminosity~\cite{Andeen:2007zc}, while the overall normalizations 
in the \zhl, \hwwlvlv, and \ssem\ analyses
are determined from 
the mass peak of $Z\rightarrow \ell\ell$ and $Z\rightarrow\tau^+\tau^-$  
decays in data
assuming the NNLO $Z/\gamma^*$ cross section, reducing the uncertainty
to about 1\%.
The \hbb~analyses have an uncertainty of 1--10\%\ due to the uncertainty on the 
$b$-tagging rate, depending on the number and quality of tagged jets.
All analyses take into account uncertainties on jet-energy scale, resolution, 
and jet
identification efficiency, for a combined uncertainty of $\approx 7\%$.
All analyses include uncertainties associated with measurement
and acceptances of leptons, which range from 1\%\ to 9\%\, depending on the final state.
The largest contribution to all analyses is from the uncertainty on the
simulated
background cross sections which are 4--30\%\, depending on the specific background process.
These values include both the uncertainty on
the theoretical cross section calculations and the uncertainties on
the higher-order correction factors. The uncertainty on the expected
multijet background in each channel 
is dominated by the statistics of the data sample
from which it is estimated. It is considered separately from the
uncertainties on 
the simulated backgrounds'
cross sections, and ranges from 10\%\ to 30\%.
All analyses take into account the uncertainties on the differential
cross sections arising from the choice of 
PDF set and QCD scale.
The $H\rightarrow W^+W^- \rightarrow \ell^{+} \nu \ell^{-} \bar{\nu}$
($\ell=e, \mu$) analyses divide the data according to jet multiplicity, and 
consider uncertainties on the contribution from GGF
that are a function of jet multiplicity. 
In addition, several analyses incorporate uncertainties that
alter differential distributions and kinematics of the dominant
backgrounds in the analyses.  These uncertainties are estimated from the variation of the final discriminant distribution 
due to generator and
background modeling uncertainties. 
Correlations between systematic sources are also carried
through in the calculations.  For example, the uncertainty on the
integrated luminosity is taken to be fully correlated between all signals and
backgrounds obtained from simulation. Hence any fluctuation in
luminosity is common to all channels for a single pseudoexperiment. 
All systematic uncertainties originating from a common source are assumed
to be fully correlated.

\section{Analysis technique validation with diboson production\label{sec:diboson}}

To validate our analyses techniques, we measure
diboson production cross sections in the $V+b\bar{b}$ and $\ell \nu \ell \nu$ 
final states. The analyses use multivariate discriminants
that utilize the same input variables as the discriminants used for the Higgs boson search,
but with one or more diboson processes acting as the signal.
The modified \whl , \zhl , and \zhv\ analyses (collectively called the $VZ$ analyses) 
treat the $WZ$ and $ZZ$ processes as signal, and the $WW$ process as a background.
The Higgs boson processes are not taken into account in this validation procedure.
Figure~\ref{fig:dibo}(a)
shows the background-subtracted data for the 
dijet invariant mass in the $VZ$ analyses,
and Fig.~\ref{fig:dibo}(b) for the
combined output of the
$VZ$ discriminant.
Similarly, the modified \hwwlvlv\ analysis uses the
$WW$ process as the signal with the $WZ$ and $ZZ$ processes as backgrounds. 
Figure~\ref{fig:dibo}(c) shows
the background-subtracted data for the output of the $WW$
discriminant. 
The $VZ$ analyses measure a $WZ+ZZ$ production cross section of $0.73 \pm 0.32$ times the SM prediction of
4.4 pb obtained with {\sc mcfm}. The significance for this measurement to be non-zero is 2.4 s.d.~with an expected
significance of 3.4 s.d. The $WW$ production cross section is measured to be $1.01\pm 0.06$ 
times the SM prediction of 11.3 pb, also based on
{\sc mcfm}. Both measurements 
confirm our ability to extract a small signal from a large background in 
the same final states, using the same analysis techniques 
as the search for the Higgs boson, providing validation of the background modeling.

\begin{figure}[htbp]
\includegraphics[width=0.45\textwidth]{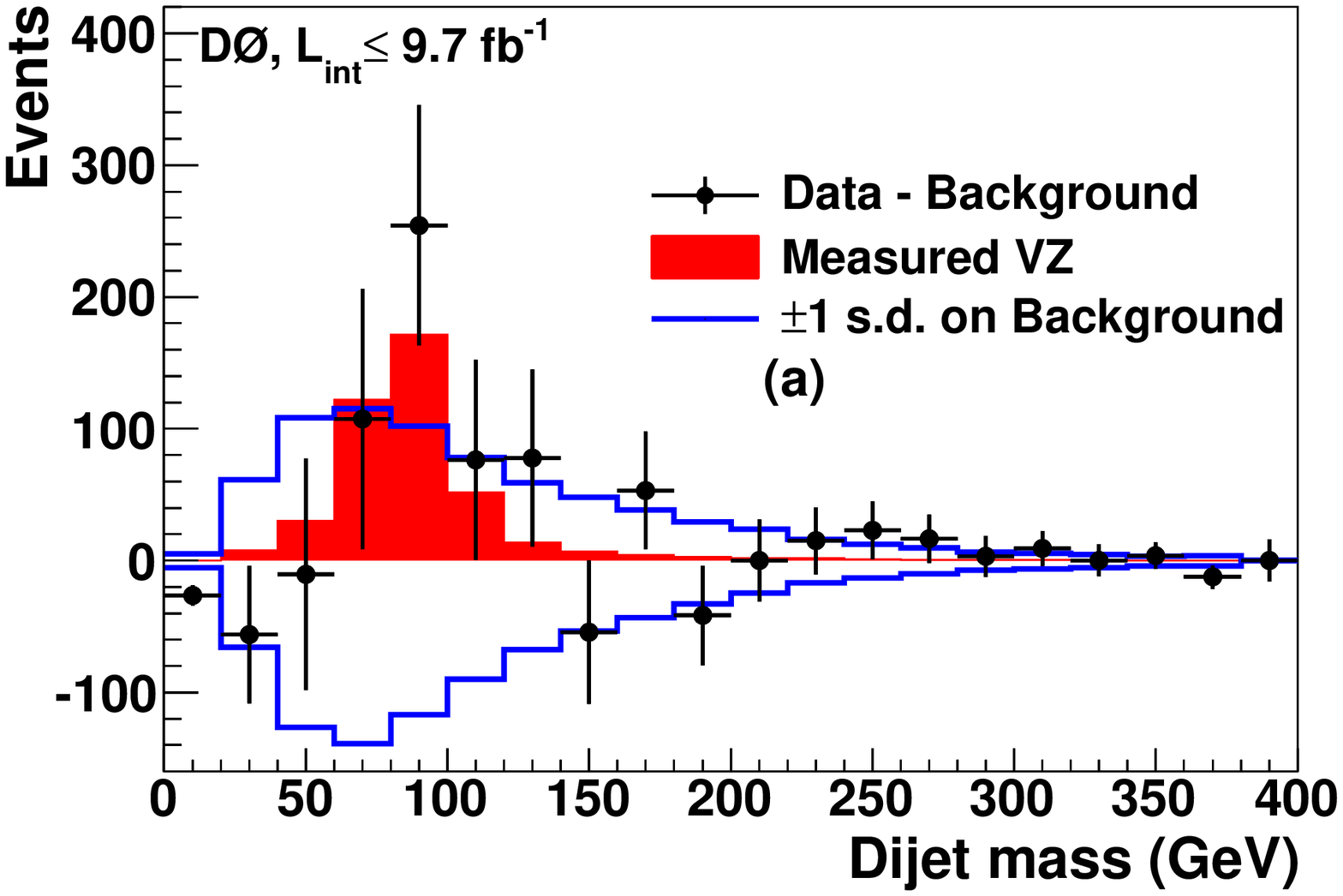}
\includegraphics[width=0.45\textwidth]{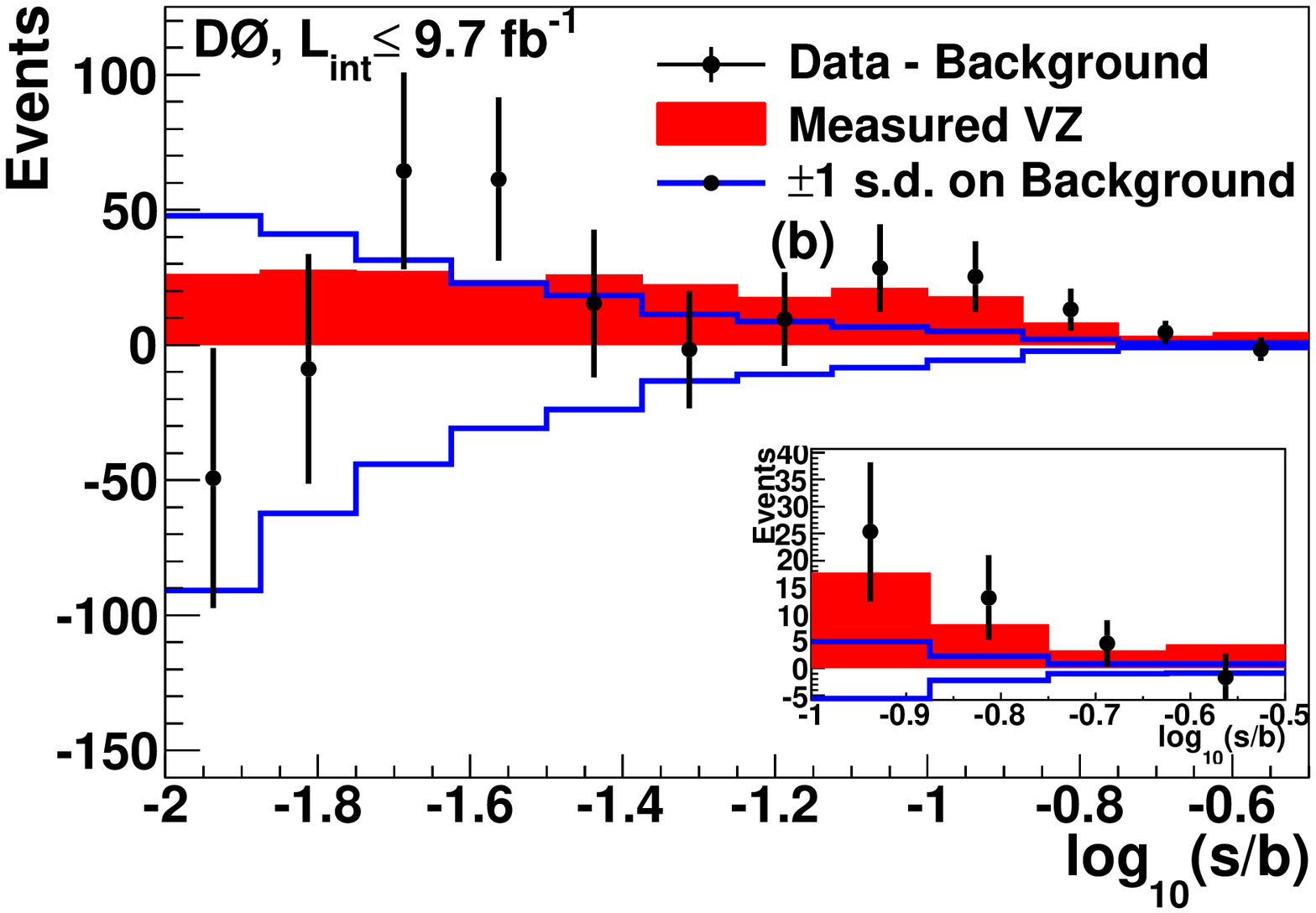}
\includegraphics[width=0.45\textwidth]{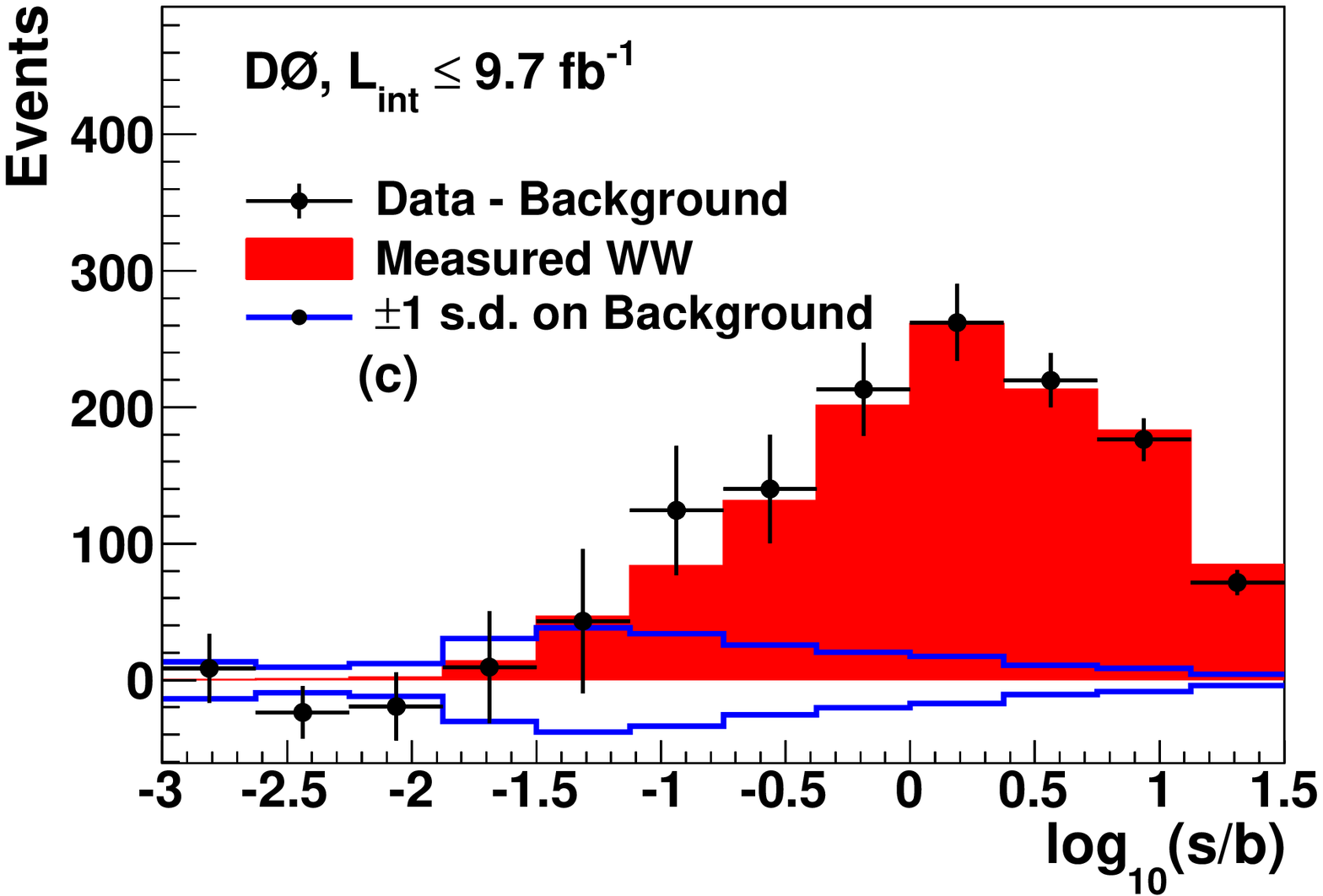}
\caption{(color online) Background-subtracted data (points with statistical errors), measured diboson signal, 
and systematic 
uncertainties after a fit to the $s+b$ hypothesis for (a) the dijet 
invariant mass in the combined $VZ\to Vb\bar{b}$ analyses, (b) the output of the
multivariate discriminant for the $VZ\to Vb\bar{b}$ analyses, 
rebinned in $\log_{10} s/b$, and (c)
the output of the multivariate discriminant for the $WW$ analysis, rebinned
in $\log_{10} s/b$.
The solid lines represent the $\pm 1$~s.d.~systematic uncertainty 
constrained by data. \label{fig:dibo}}
\end{figure}

\section{Higgs boson results\label{sec:results}}

\subsection{Limits on standard model Higgs boson production\label{sec:sm}}

We obtain limits on the product of the Higgs boson production cross section,
$\sigma_H$,
and branching fractions
$\mathcal{B}(H\rightarrow b\bar{b}/ W^{+}W^{-}/ \tau^{+}\tau^{-}/\gamma\gamma$)
using individual channels~\citeall.
We present results in terms of the ratio of the upper limit on 
$\sigma_H$ at 95\%\ C.L.~relative to the SM predicted values as a function of $M_H$,
where the relative cross sections and branching fractions are kept as predicted by the SM.
The SM prediction is therefore excluded at the 95\%\ C.L.~for the $M_H$ values at which the ratio falls below unity.

The LLR distributions for the full combination
are shown in Fig.~\ref{fig:allLLR}. Included in these figures are the
median LLR values expected for the $s+b$ hypothesis
($\text{LLR}_{s+b}$), $b$ hypothesis ($\text{LLR}_{b}$), and the results
observed in data ($\text{LLR}_ {\text{obs}}$). The shaded bands represent the $\pm1$ 
and $\pm2~\text{s.d}.$ departures for $\text{LLR}_{b}$. These
distributions can be interpreted as follows:

\begin{enumerate}[(i)]
\item The separation between $\text{LLR}_{b}$ and $\text{LLR}_{s+b}$ provides 
a measure of the discriminating power of the search,  and illustrates 
the effectiveness of the analysis to separate the $s+b$ 
and $b$ hypotheses.

\item The width of the $\text{LLR}_{b}$ distribution (shown here as $\pm1$ and
$\pm2$~s.d.~bands) provides an estimate of the
sensitivity of the analysis to a signal-like background fluctuation
in the data, taking the systematic
uncertainties into account.  For example, the sensitivity is limited when
a 1~s.d.~background fluctuation is large compared to the difference between
the $s+b$ and $b$ expectations.

\item The value of $\text{LLR}_ {\text{obs}}$ relative to $\text{LLR}_{s+b}$ and $\text{LLR}_{b}$
indicates whether the data distribution appears to be more
$s+b$-like or $b$-like.  The
significance of any departures of $\text{LLR}_ {\text{obs}}$ from $\text{LLR}_{b}$ can be
evaluated through the width of the $\text{LLR}_{b}$ distribution.

\end{enumerate}

As shown in Table~\ref{tab:chans}, only the \whl\ and \zhl\ channels contribute to the combination below $M_H=100$~GeV.
Figure~\ref{fig:allLLR} shows that the observed LLR is compatible with the $s+b$ hypothesisfor $120 < M_H < 145$ GeV.

Figure~\ref{fig:allHI} shows the expected and observed upper limits
on $\sigma_H$ at 95\%\ C.L.~relative
to the SM, for the
mass region $90 \leq M_H \leq 200$~GeV, for all analyses combined. 
These results are also summarized in
Table~\ref{tab:limits}. We exclude the SM Higgs boson at 95\%\ C.L. 
in the mass ranges $90 < M_H < \excllow$~GeV 
and $\exclmin <M_H< \exclmax$~GeV. Our 
expected exclusion range is
$\exclminexp <M_H<\exclmaxexp$~GeV.

\begin{figure}[htp]
\begin{centering}
\includegraphics[width=0.45\textwidth]{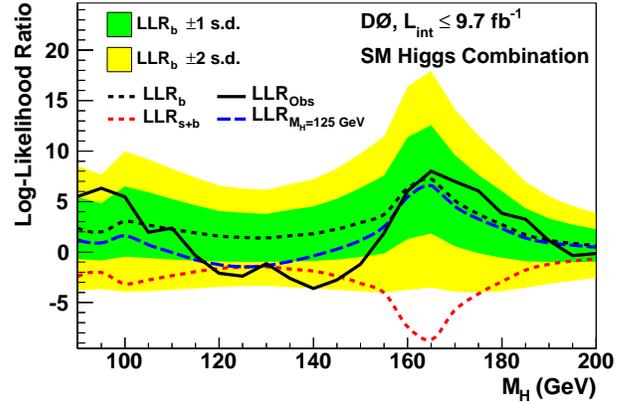}
\caption{
\label{fig:allLLR} (color online)
The observed (black solid line) and expected
LLRs for the $b$ (black short-dashed line) and $s+b$
hypotheses (red/light short-dashed line), as well as
the LLR expected in the presence of a SM Higgs boson with
$M_H = 125$~GeV (blue long-dashed line)
for all analyses combined for the range 
$90 \leq M_H \leq 200$~GeV. 
The shaded bands correspond,
respectively, to the
regions enclosing $\pm1$ and $\pm2$~s.d.~fluctuations of the background.
}
\end{centering}
\end{figure}

\begin{figure}[htbp]
\begin{centering}
\includegraphics[width=0.45\textwidth]{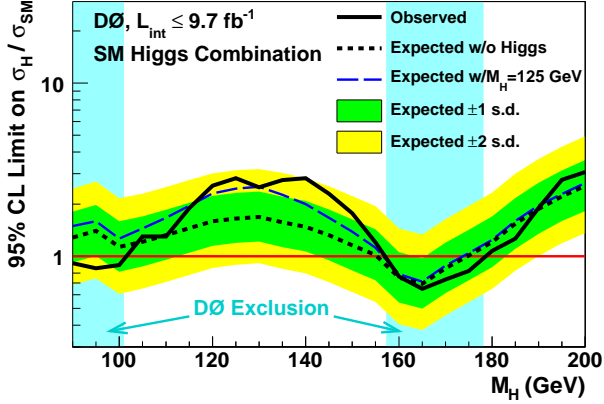}
\caption{
\label{fig:allHI} (color online)
Expected (median) and observed 
ratios for the upper limits of the cross section 
$\sigma_H$ at
95\%~C.L.~relative to the SM values
for all analyses combined
for the range $90 \leq M_H \leq 200$~GeV.  The shaded 
bands correspond to the regions enclosing $\pm1$ and $\pm2$~s.d.~fluctuations of the background, respectively. The long-dashed
line represents the expectation if a $M_H=125$ GeV Higgs boson were 
present in the data with the SM cross section.}
\end{centering}
\end{figure}

\begin{table*}[tp]
\caption{
Expected (median) and observed
upper limits on the cross sections
relative to the SM at 95\%\ C.L.~for the combined 
  analyses for the range $90 \leq M_H \leq 200$~GeV.
\label{tab:limits}}
\begin{ruledtabular}
\begin{tabular}{lccccccccccccccccccccccc}
$\! M_{H}\!$(GeV) & 90 & 95 &100 &105 &110 &115 &120 &125 &130 &135 &140 &145 &150 &155 &160 &165 &170 &175 &180 &185 &190 &195 &200 \\
\hline
$\!$Expected &1.29 &1.40 &1.13 &1.21 &1.32 &1.45 &1.59 &1.66 &1.69 &1.58 &1.49 &1.33 &1.17 &1.02 &0.75 &0.70 &0.86 &1.02 &1.21 &1.55 &1.89 &2.22 &2.55 \\
$\!$Observed &0.96 &0.89 &0.95 &1.39 &1.39 &1.99 &2.66 &2.92 &2.56 &2.79 &2.88 &2.36 &1.84 &1.23 &0.78 &0.66 &0.75 &0.85 &1.11 &1.31 &1.96 &2.85 &3.12 \\

\end{tabular}
\end{ruledtabular}
\end{table*}

Figure~\ref{fig:allCLSB} shows the values for the observed $CL_{s+b}$ and its
expected behavior as a function of $M_H$. The quantity $CL_{s+b}$
corresponds to the $p$-value for the $s+b$ hypothesis. 
Figure~\ref{fig:allCLB} shows the
quantity $1-CL_{b}$, which is the $p$-value for the $b$
hypothesis.  These probabilities are local $p$-values, corresponding
to searches for each value of $M_H$ separately.
These two $p$-values ($CL_{s+b}$ and $1-CL_{b}$) provide information about the
consistency of their respective hypotheses with the observed data
at each value of $M_H$.
Small values indicate rejection of the hypothesis and values 
above 50\%
indicate general agreement between the hypothesis in question
and the data.  As can be seen in Fig.~\ref{fig:allCLSB}, the
observed value of $CL_{s+b}$ drops to $\approx 1\%$ for 
$M_H=160$~GeV, indicating limited consistency with the
$s+b$ hypothesis around this mass. In contrast, the
observed value of $CL_{s+b}$ is close to unity for $120 \le M_{H}
\le 145$~GeV, whereas $1-CL_{b}$ is small.
At $M_{H}=125~(140)$~GeV, the value of $1-CL_{b}$ is
$\ABElocalpval$ ($\ADZlocalpval$),
corresponding to
$\ABElocalzval$ ($\ADZlocalzval$)~s.d.~above the background prediction.

\begin{figure}[htbp]
\begin{centering}
\includegraphics[width=0.45\textwidth]{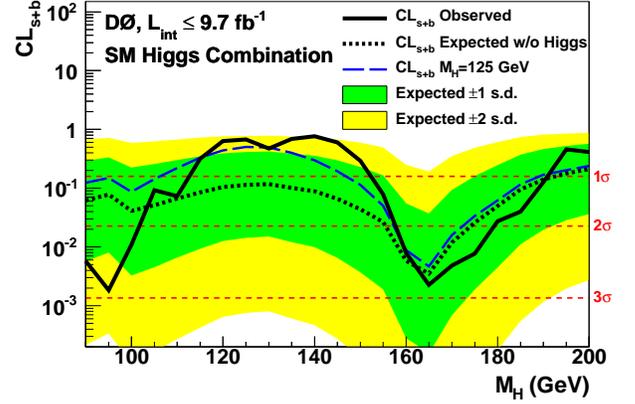}
\caption{
\label{fig:allCLSB} (color online)
The observed (black solid line) and expected
 $CL_{s+b}$ ($s+b$ $p$-value)
for the no-Higgs boson hypothesis (black short-dashed line) and 
in the presence of a SM Higgs boson with
$M_H = 125$~GeV (blue long-dashed line)
for all analyses combined for the range 
$90 \leq M_H \leq 200$~GeV. 
The shaded bands correspond,
respectively, to the
regions enclosing $\pm1$ and $\pm2$~s.d.~fluctuations of the background.
The three red
horizontal dashed lines indicate the $p$-values corresponding to
significances of 1, 2 and 3 s.d.}
\end{centering}
\end{figure}

\begin{figure}[htbp]
\begin{centering}
\includegraphics[width=0.45\textwidth]{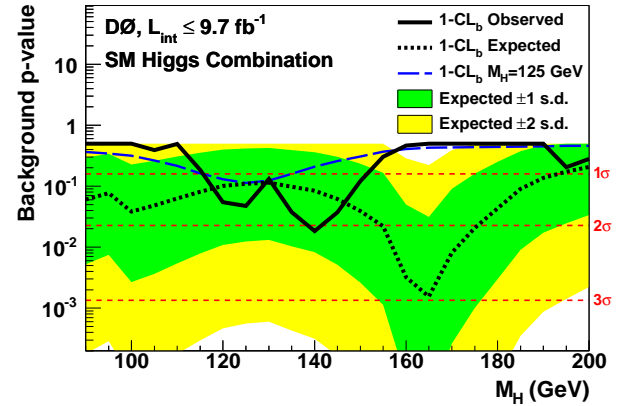}
\caption{
\label{fig:allCLB} (color online)
The observed (black solid line) and expected (black short-dashed line) 
$1-CL_{b}$ (background $p$-value)
for all analyses combined for the range 
$90 \leq M_H \leq 200$~GeV. 
Also shown
is the expected background $p$-value for a presence of a $M_H=125$~GeV 
SM Higgs boson signal
in the data (blue long-dashed line).
The shaded bands correspond,
respectively, to the
regions enclosing $\pm1$ and $\pm2$~s.d.~fluctuations of the background.
The three red
horizontal dashed lines indicate the $p$-values corresponding to
significances of 1, 2 and 3~s.d.
}
\end{centering}
\end{figure}

\begin{figure}[htbp]
\begin{centering}
\includegraphics[width=0.45\textwidth]{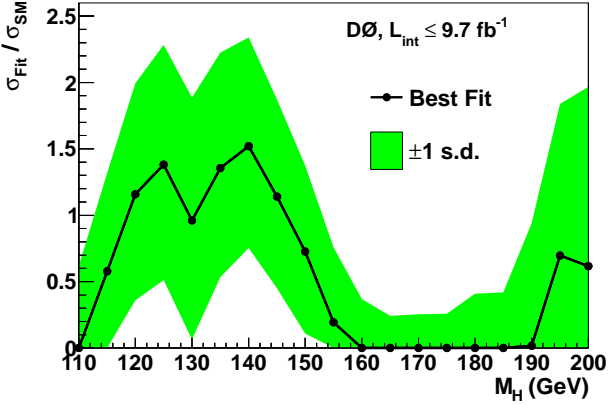}
\caption{
\label{fig:comboXsec}  (color online)
The best fit of the ratio $\sigma_H/(\sigma_{H})_{\rm SM}$
as a function of $M_H$
for all analyses
combined for the $110 \leq M_H \leq 200$~GeV. This 
indicates the values of the Higgs boson cross section that best
match the observed data.
The light shaded band indicates the $\pm1$~s.d.~region
departure from the fit. The fit result is zero for masses below 110~GeV.}
\end{centering}
\end{figure}

As a further investigation of this excess,
we present in Fig.~\ref{fig:comboXsec} 
the best fit of the data to the ratio of $\sigma_H$
to the SM
prediction ($\sigma_{\text{Fit}}/\sigma_{\text{SM}}$).  
The result of this fit, shown along
with its band of $\pm1~$s.d., yields a
signal rate of approximately a factor of 1.4 larger than
the SM 
cross section for $M_H$ between 120~GeV and 145~GeV.
For  $M_H=125$~GeV, we obtain a ratio of $1.4 \pm 0.9$.
The associated production analyses with $H \to b\bar{b}$
decay and the \hwwlvlv\ analyses dominate our sensitivity.
 The dijet invariant mass resolution is approximately
15\%\ for associated production with $H \to b\bar{b}$ decay. 
The mass resolution for the analyses with $H\to W^+W^-$ decay is 
poor due to the undetected neutrinos in the final state.
We therefore expect a Higgs boson signal to appear as a broad
excess over background, rather than a narrow resonance such as
that expected at the LHC in the $H\to \gamma\gamma$ or $H\to ZZ \to 4\ell$ 
final states.

We study the excess at low mass by separating the major contributing
sources according to the Higgs boson decay: $H\rightarrow b\bar{b}$, 
$H\rightarrow W^+W^-$, $H\rightarrow\tau^+\tau^-$ and 
$H\rightarrow\gamma\gamma$ final states.  
Figure~\ref{fig:hbbLLR} 
shows the LLR values from the combination of the results from
the $ZH\rightarrow \ell\ell b\bar{b}$, $ZH\rightarrow
\nu\nu b\bar{b}$ and $WH\rightarrow \ell\nu b\bar{b}$ searches, and
illustrates a small excess 
that is compatible with the 
SM Higgs boson 
expected rate for $120 \leq M_{H} \leq 145$~GeV.  Figure~\ref{fig:hwwLLR}
shows the LLR values from the combination of the results
from searches for
$H\rightarrow W^+W^- \rightarrow
\ell\nu\ell\nu$, $H\rightarrow W^+W^- \rightarrow \ell\nu jj$, and $VH
\rightarrow VWW$, together with the $WW$-dominated 
subchannels from the \tautaujj\ analysis,
and shows a similar excess of data over the background for $110 \leq M_H \leq 150$~GeV. At higher
masses, where the Tevatron sensitivity to Higgs boson production is the largest, 
the LLR favors the $b$ hypothesis.
Figure \ref{fig:htautauLLR} 
shows the LLR values from 
the combination of the 
$\tau\tau$-dominated \tautaujj\ subchannels and the \ttm\ 
analysis, in which a significant fraction of the Higgs boson decays are to 
$\tau^+\tau^-$ pairs.
Figures~\ref{fig:hbbLimit}--\ref{fig:htautauLimit}, as well as Tables
\ref{tab:HBBlimits}--\ref{tab:Htautaulimits}, show the
expected and observed 95\%\ C.L.~cross section limits in terms of ratio to the
SM predictions

for $H\rightarrow b\bar{b}$, 
$H\rightarrow W^+W^-$, and $H\rightarrow \tau^+\tau^-$ final
states, respectively. The corresponding figures for the $H\rightarrow\gamma\gamma$ analysis can be found in Ref.~\cite{Abazov:hgg}.
Figure~\ref{fig:sigstr_sep} shows the
best fit of the ratio $\sigma_H\cdot\mathcal{B}/(\sigma_{H}\cdot\mathcal{B})_{SM}$ for $M_H=125$~GeV 
in each of the Higgs boson decay channels considered, as well as the central value 
for all analyses combined. These values are also given in Table~\ref{tab:sigstr_sep}.

\begin{figure}[htp]
\begin{centering}
\includegraphics[width=0.45\textwidth]{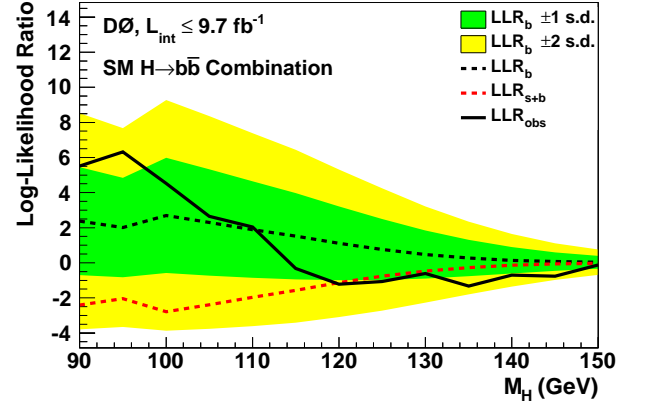}
\caption{(color online) 
The observed (black solid line) and expected
LLRs for the $b$ (black short-dashed line) and $s+b$
hypotheses (red/light short-dashed line)
for the combined $WH/ZH,H\rightarrow b\bar{b}$ analyses
for the range 
$90 \leq M_H \leq 150$~GeV. 
The shaded bands correspond,
respectively, to the
regions enclosing $\pm1$ and $\pm2$~s.d.~fluctuations of the background.
\label{fig:hbbLLR}}
\end{centering}
\end{figure}

\begin{figure}[htbp]
\begin{centering}
\includegraphics[width=0.45\textwidth]{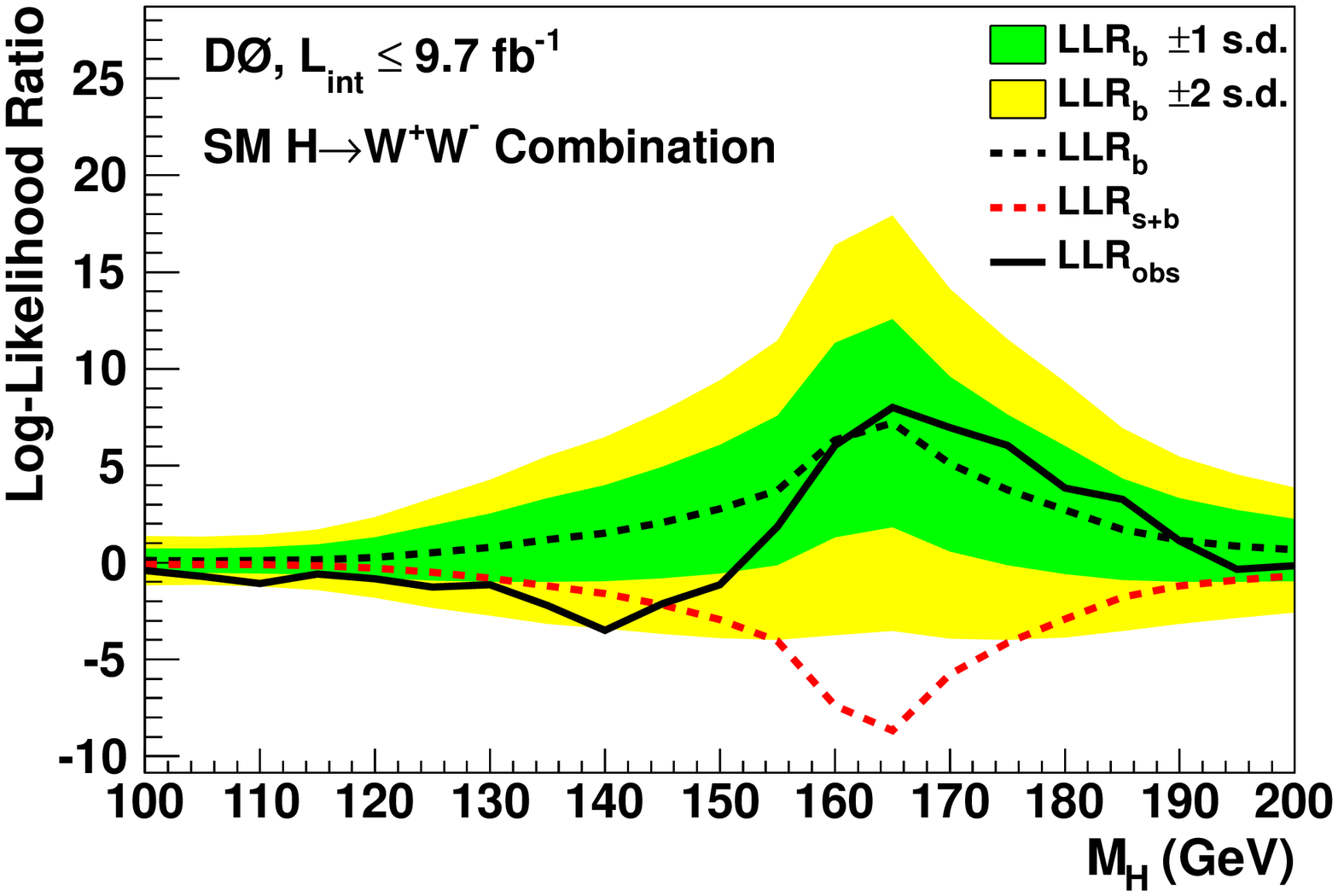}
\caption{(color online)
The observed (black solid line) and expected
LLRs for the $b$ (black short-dashed line) and $s+b$
hypotheses (red/light short-dashed line)
for the combined $WH/ZH/H, H\rightarrow W^+W^-$ analyses
for the range 
$100 \leq M_H \leq 200$~GeV. 
The shaded bands correspond,
respectively, to the
regions enclosing $\pm1$ and $\pm2$~s.d.~fluctuations of the background.
\label{fig:hwwLLR}}
\end{centering}
\end{figure}

\begin{figure}[htbp]
\begin{centering}
\includegraphics[width=0.45\textwidth]{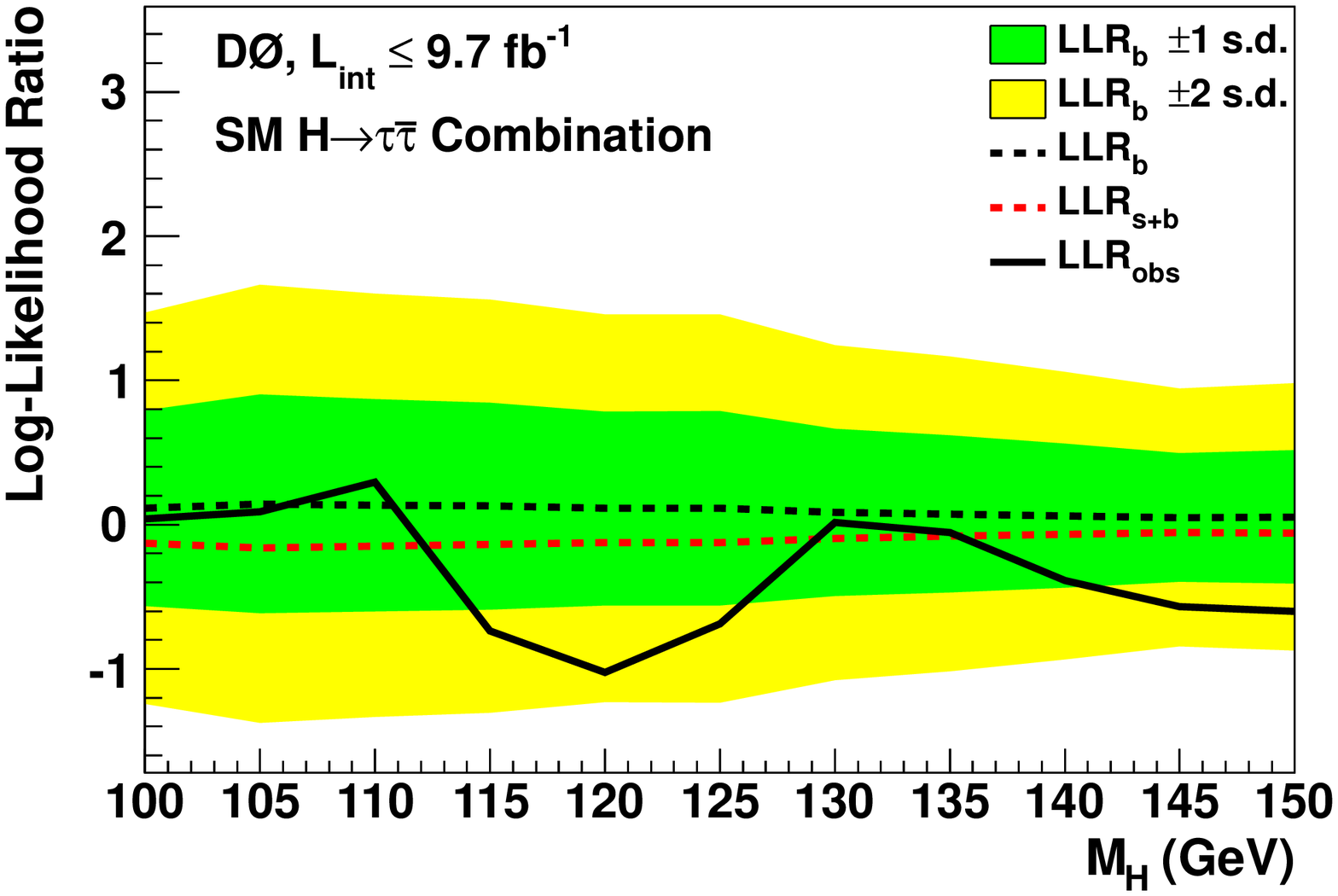}
\caption{(color online) 
The observed (black solid line) and expected
LLRs for the $b$ (black short-dashed line) and $s+b$
hypotheses (red/light short-dashed line)
for the combined \ttm\ and 
\tautaujj\ analyses
for the range 
$100 \leq M_H \leq 150$~GeV. 
The shaded bands correspond,
respectively, to the
regions enclosing $\pm1$ and $\pm2$~s.d.~fluctuations of the background.
\label{fig:htautauLLR}}
\end{centering}
\end{figure}

\begin{figure}[htbp]
\begin{centering}
\includegraphics[width=0.45\textwidth]{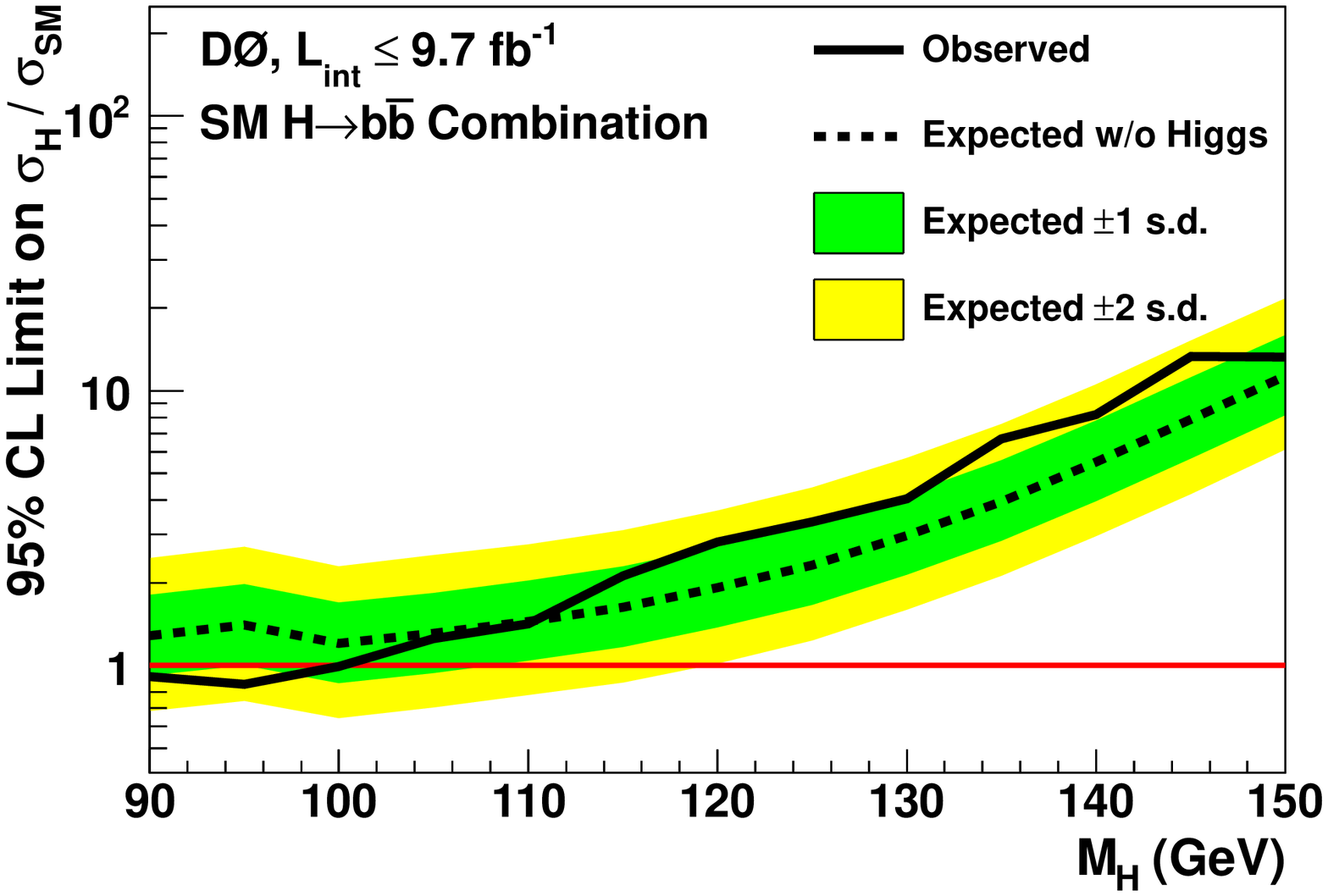}
\caption{
\label{fig:hbbLimit} (color online)
Expected (median) and observed 
ratios for the upper limits of the cross section 
$\sigma_H$ at
95\%~C.L.~relative to the SM values
for the combined $WH/ZH, H\rightarrow b\bar{b}$ analyses
for the range $90 \leq M_H \leq 150$~GeV.  The shaded 
bands correspond to the regions enclosing $\pm1$ and $\pm2$~s.d.~fluctuations of the background, respectively. 
}
\end{centering}
\end{figure}

\begin{figure}[htbp]
\begin{centering}
\includegraphics[width=0.45\textwidth]{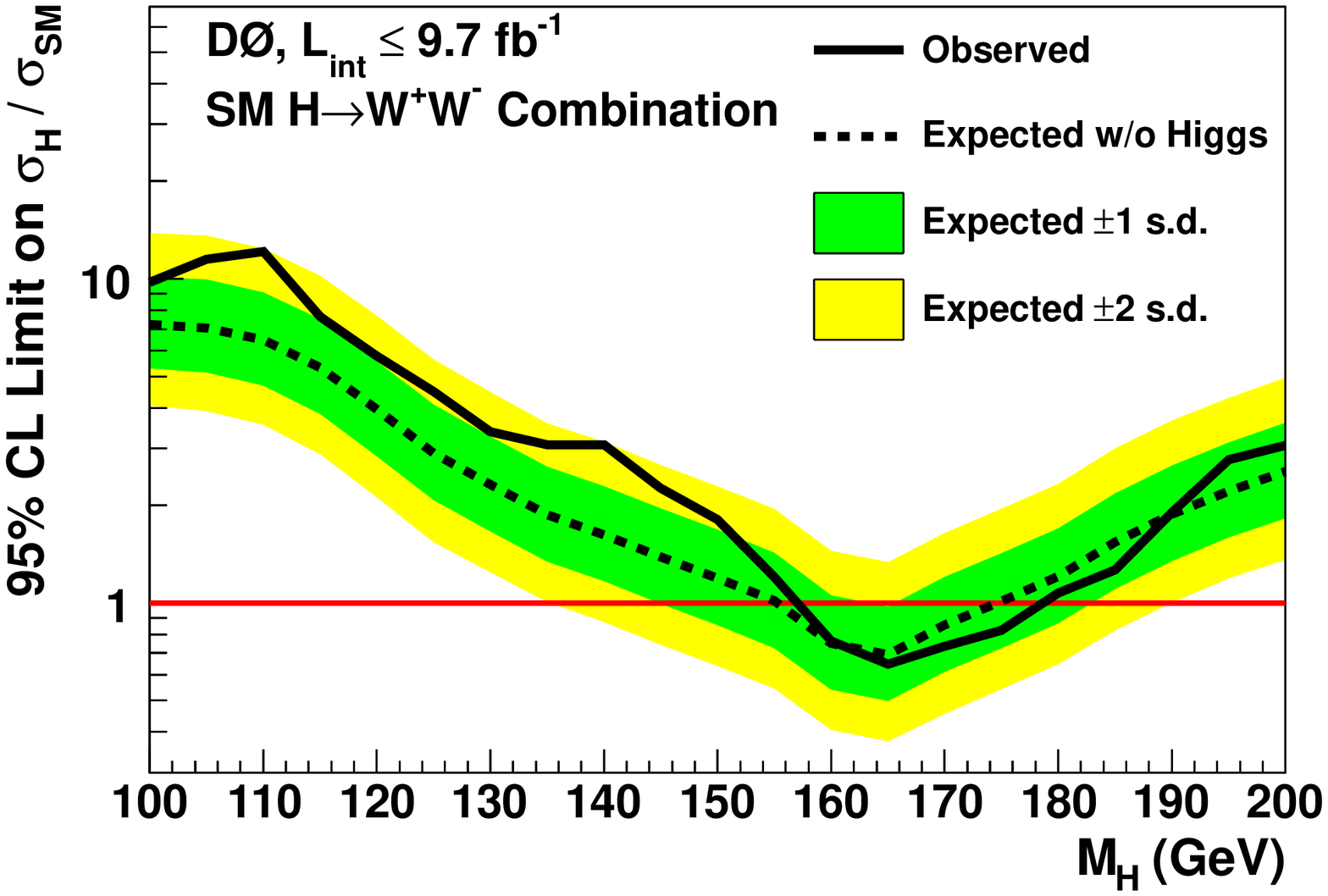}
\caption{
\label{fig:hwwLimit} (color online)
Expected (median) and observed 
ratios for the upper limits of the cross section 
$\sigma_H$ at
95\%~C.L.~relative to the SM values
for the combined $WH/ZH/H, H\rightarrow W^+W^-$ analyses
for the range $100 \leq M_H \leq 200$~GeV.  The shaded 
bands correspond to the regions enclosing $\pm1$ and $\pm2$~s.d.~fluctuations of the background, respectively. 
}
\end{centering}
\end{figure}

\begin{figure}[htbp]
\begin{centering}
\includegraphics[width=0.45\textwidth]{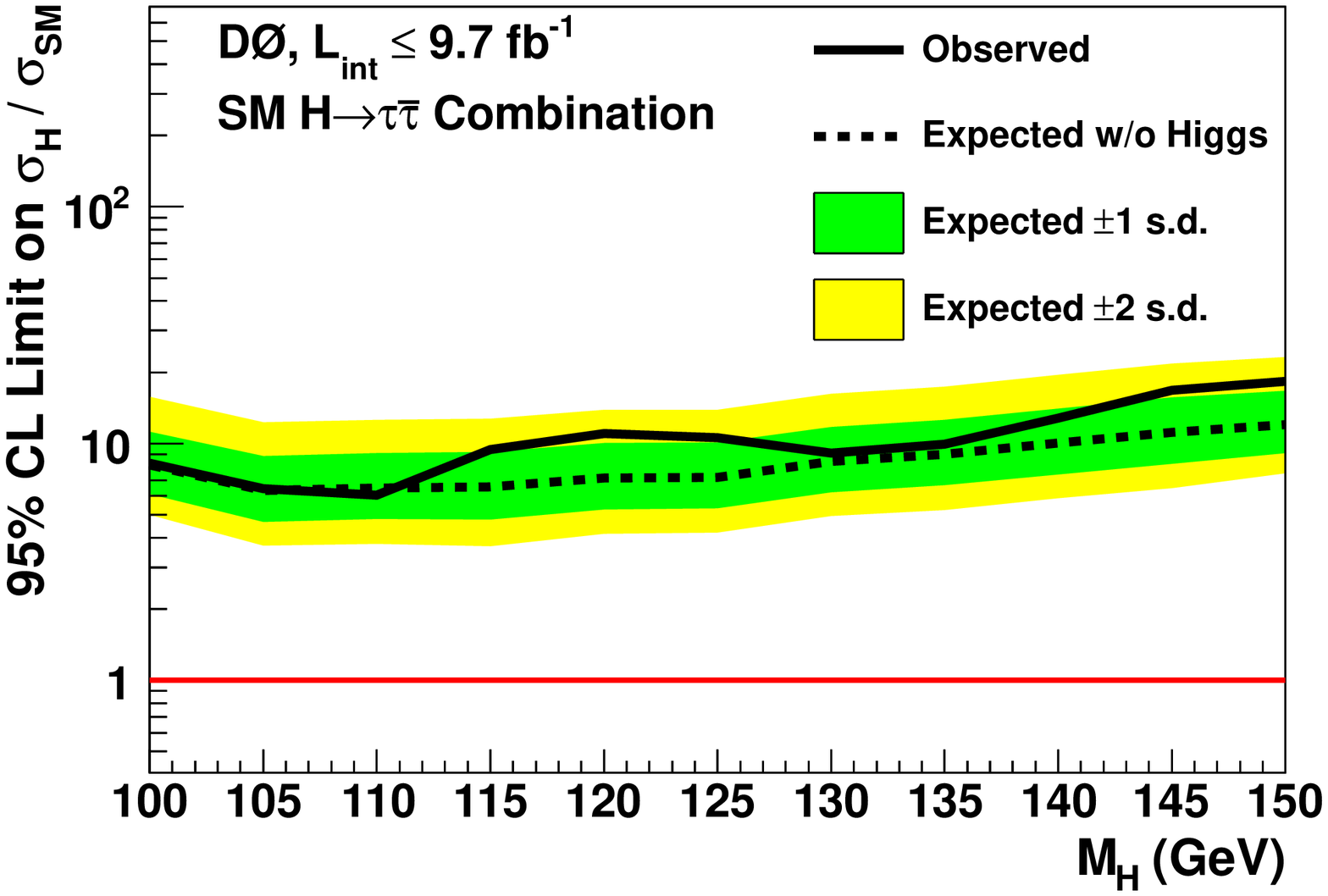}
\caption{
\label{fig:htautauLimit} (color online)
Expected (median) and observed 
ratios for the upper limits of the cross section 
$\sigma_H$ at
95\%~C.L.~relative to the SM values
for the combined \ttm\ and \tautaujj\ analyses 
for the range $100 \leq M_H \leq 150$~GeV.  The shaded 
bands correspond to the regions enclosing $\pm1$ and $\pm2$~s.d.~fluctuations of the background, respectively. 
}
\end{centering}
\end{figure}

\begin{figure}[htbp]
\begin{centering}
\includegraphics[width=0.45\textwidth]{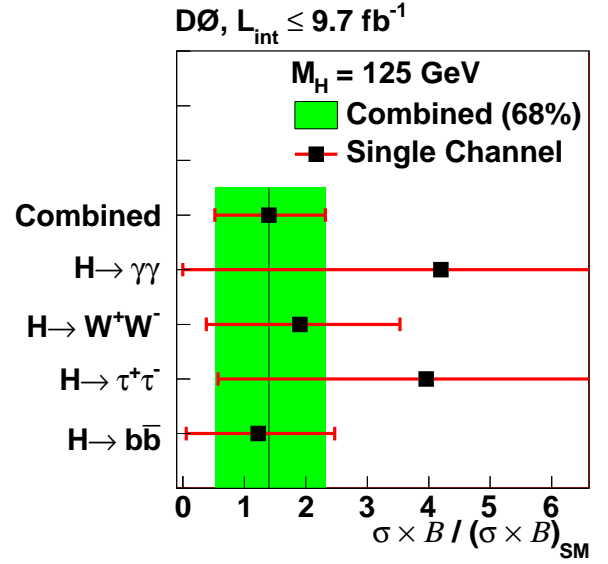}
\caption{
\label{fig:sigstr_sep} (color online)
The best fit of 
$\sigma_{H}\cdot\mathcal{B}/(\sigma_{H}\cdot\mathcal{B})_{\rm SM}$
for various Higgs boson decays for
$M_H=125$~GeV.
The central value for all combined analyses is shown 
with its 1 s.d.~band (shaded area).
}
\end{centering}
\end{figure}

\begin{table*}[htbp]
\caption{Expected (median) and observed
upper limits for $\sigma \times \mathcal{B}(H\rightarrow b\bar{b})$ 
relative to the SM at 95\%\ C.L.~for the 
combined $WH/ZH, H\rightarrow b\bar{b}$
  analyses for the range $90 \leq M_H \leq 150$~GeV.
\label{tab:HBBlimits}}
\begin{ruledtabular}
\begin{tabular}{lccccccccccccc}
$M_{H}$ (GeV) & 90 & 95 &100 &105 &110 &115 &120 &125 &130 &135 &140 &145 &150 \\
\hline
Expected & 1.29 &1.40 &1.21 &1.31 &1.45 &1.63 &1.92 &2.33 &2.99 &3.96 &5.52 &7.91 &11.35 \\
Observed & 0.96 &0.89 &1.05 &1.33 &1.51 &2.25 &2.96 &3.49 &4.29 &6.92 &8.65 &13.85 &13.90 \\

\end{tabular}
\end{ruledtabular}
\end{table*}

\begin{table*}[htbp]
\caption{
Expected (median) and observed
upper limits for $\sigma \times \mathcal{B}(H\rightarrow W^+W^-)$ 
relative to the SM at 95\%\ C.L.~for the 
combined $WH/ZH/H, H\rightarrow W^+W^-$
  analyses for the range $100 \leq M_H \leq 200$~GeV.
\label{tab:HWWlimits}}
\begin{ruledtabular}
\begin{tabular}{lccccccccccccccccccccc}
$M_{H}$ (GeV) &100  &105 &110 &115 &120 &125 &130 &135 &140 &145 &150 &155 &160 &165 &170 &175 &180 &185 &190 &195 &200 \\
\hline
Expected &7.25 &7.09 &6.49 &5.34 &3.97 &2.92 &2.33 &1.88 &1.64 &1.40 &1.20 &1.02 &0.75 &0.70 &0.86 &1.02 &1.21 &1.55 &1.89 &2.22 &2.55 \\
Observed &9.98 &11.69 &12.38 &7.70 &5.84 &4.55 &3.42 &3.15 &3.14 &2.30 &1.86 &1.23 &0.78 &0.66 &0.75 &0.85 &1.11 &1.31 &1.96 &2.85 &3.12\\

\end{tabular}
\end{ruledtabular}
\end{table*}

\begin{table*}[htbp]
\caption{
Expected (median) and observed
upper limits for $\sigma \times \mathcal{B}(H\rightarrow \tau^+\tau^-)$ 
relative to the SM at 95\%\ C.L.~for the 
combined \ttm\ and \tautaujj\
  analyses for the range $100 \leq M_H \leq 150$~GeV.
\label{tab:Htautaulimits}}
\begin{ruledtabular}
\begin{tabular}{lccccccccccccccccccccc}
$M_{H}$ (GeV) &100  &105 &110 &115 &120 &125 &130 &135 &140 &145 &150 \\
\hline
Expected &8.22 &6.39 &6.54 &6.59 &7.21 &7.25 &8.46 &9.05 &10.11 &11.28 &12.11 \\
Observed &8.42 &6.64 &6.20 &9.70 &11.29 &10.84 &9.35 &10.17 &13.07 &17.16 &18.59\\

\end{tabular}
\end{ruledtabular}
\end{table*}

\begin{table}[htbp]
\caption{The best fit Higgs boson cross section times branching fraction as a ratio to the SM cross section times branching fraction for $M_H=125$ GeV
for the individual combinations according to Higgs boson decay mode, as well as the full combination.}
\begin{tabular}{lr}
\hline
\hline
\vspace{1mm}
Combined & $1.40^{+0.92}_{-0.88}$ \T \\ \hline
\vspace{1mm}
$H\rightarrow \gamma\gamma$ & $4.20^{+4.60}_{-4.20}$ \T \\
\vspace{1mm}
$H\rightarrow W^+W^-$ & $1.90^{+1.63}_{-1.52}$ \T \\
\vspace{1mm}
$H\rightarrow \tau^+\tau^-$ & $3.96^{+4.11}_{-3.38}$ \T \\
\vspace{1mm}
$H\rightarrow b\bar{b}$ & $1.23^{+1.24}_{-1.17}$ \T \\
\hline
\hline
\vspace{1mm}
\end{tabular}
\label{tab:sigstr_sep}
\end{table}

\subsection{Interpretation in fourth generation and Fermiophobic Higgs boson models\label{sec:bsm}}

We also interpret our Higgs boson searches in
models containing a fourth generation of fermions, and models with a fermiophobic 
Higgs boson. The fourth generation models~\cite{Holdom:2009rf} feature a modified $Hgg$ coupling, leading to 
a nearly order of magnitude enhancement in the GGF cross section
relative to the SM~\cite{Arik:2005ed, Kribs:2007nz,Anastasiou:2010bt}. Previous
interpretations of SM Higgs boson searches within the
context of a fourth generation of fermions at the Fermilab Tevatron Collider
exclude $131 < M_{H} < 207$~GeV~\cite{Aaltonen:2010sv}. Both 
ATLAS~\cite{Aad:2011qi} and CMS~\cite{CMS-SM4}
have performed similar searches, which exclude,
respectively, $140 < M_{H} < 185$~GeV
and $110 < M_{H} < 600$~GeV.
Although the larger coupling increases the decay 
width to $gg$, the $WW^{*}$ decay mode 
remains dominant for $M_{H} > 135$~GeV.
There is also a small contribution
from $H\rightarrow ZZ^*\rightarrow \ell\ell\nu\nu$ production that increases with $M_H$.
We consider two fourth generation 
scenarios: (i) a ``low mass'' scenario in which the mass of 
the fourth generation neutrino is set to $m_{\nu 4}=80$~GeV, 
and the mass of the 
fourth generation charged lepton $m_{\ell4}$ is set to 100~GeV,
and (ii) a ``high mass'' scenario in which 
$m_{\nu4} = m_{\ell4} = 1$~TeV, so that 
the fourth generation leptons do not affect the decay branching fractions of the Higgs boson. In both scenarios 
the fourth generation quark masses are set 
to be those of the high mass scenario in Ref.~\cite{Anastasiou:2010bt}. 

We consider only \pggh\ production 
and the \hwwlvlv\ and \hwwlnuqq$(qq)$  
channels to set limits on the fourth generation models, and
also set a limit on
$\sigma(gg \to H)\times \mathcal{B}(H \to W^{+}W^{-})$.
We scale the product of the cross sections 
and branching fractions to the results from {\sc hdecay}, modified 
to include the fourth generation. 
We retrain our multivariate discriminants to take only the above signals 
into account, and do not include events with two or more jets 
in the \hwweemm\ analyses. We also do not include the theoretical 
uncertainty on $\sigma(gg \to H)\times \mathcal{B}(H \to W^{+}W^{-})$
since the absolute cross section limits do not depend on the prediction. We include the 
theoretical uncertainties for limits on ratios
to cross sections.

Figure \ref{fig:SM4} shows the combined limits on $\sigma(gg \to H)\times \mathcal{B}(H \to W^{+}W^{-})$,
along with the fourth generation theory predictions for the high mass 
and low mass scenarios.
We exclude a SM-like Higgs boson in the range $125 < M_{H} < 218$~GeV 
at  95\%\ C.L., with an expected exclusion 
range of $122<M_{H}<232$~GeV in the low mass scenario. In the high mass 
scenario, the observed (expected)
exclusion range is $125 < M_{H} < 228$ ($122 < M_{H} < 251$)~GeV.

In the fermiophobic model (FHM), the lightest Higgs boson $H_f$ couplings to fermions
vanish at leading order,
but otherwise $H_f$ is like the SM Higgs boson. 
Hence, $gg\to H_f$ production is negligible, and $H_f$ decays to 
fermions are forbidden, but $V+H_f$ and vector boson fusion 
\pvbf\
production remain nearly unchanged relative to the SM. 
The $WW$, $ZZ$, $\gamma\gamma$, and $Z\gamma$
decays comprise nearly the entire decay width. 
For all $M_{H_f}$ the $H_f\to W^{+}W^{-}$ decay has the 
largest branching fraction. The $H_f\to\gamma\gamma$ branching fraction 
is greatly enhanced over the SM for all $M_{H_f}$, and
it provides most of the search sensitivity for $M_{H_f}< 120$~GeV.

The CDF and D0 Collaborations have previously published results in the $H_{f}\to\gamma\gamma$
decay channel~\cite{Collaboration:2012pa,Abazov:2011ix}. 
The analyses described here supersede previous FHM searches at D0.
The ATLAS and CMS Collaborations have performed fermiophobic searches, 
and exclude $110 < M_{H_f} < 118.0$~GeV,
$119.5 < M_{H_f} < 121.0$~GeV~\cite{Aad:2012yq}, and
$110 < M_{H_f} < 147$~GeV~\cite{CMS-SM4} using $\gamma\gamma$ final 
states, and
$110 < M_{H_f} < 194$~GeV when other final states are included~\cite{CMS:2012bd}.

We combine the 
$H\to\gamma\gamma$ and $H\to W^+W^-$ decay channels, produced either in association
with a $V$ boson, or in VBF, for the FHM interpretation. 
We reoptimize the SM \hgg\ analysis to take into account the 
different kinematics in the FHM, 
e.g., the presence of an associated vector boson in the FHM,
or recoiling quark jets in VBF, which
shift the transverse momentum spectrum of the Higgs boson to
higher values than in the SM.
Likewise, we retrain the  multivariate discriminants for the \hwwlnulnu\ analyses
to account for the suppressed GGF process in the FHM.
We retain the existing subdivision into categories that are based on the 
number of reconstructed jets in the event.
The other SM $H\to WW$ analyses can be interpreted directly in the FHM without 
reoptimization,
after separating the relative contributions from GGF, $WH$, $ZH$, and 
VBF in each contributing channel, removing the GGF component,
and scaling the remaining signal contributions by the ratio of the branching 
fraction in the FHM and SM, 
$\mathcal{B}(H_f \to WW)/\mathcal{B}(H_{SM} \to WW)$. 
Figure~\ref{fig:FHM} shows the combined FHM limits. The observed (expected) 95\%\ C.L.~exclusion range is $100 < M_{H_f} < 114$ ($100 < M_{H_f} < 117$)~GeV.

\begin{figure}[hbtp]
\includegraphics[width=0.45\textwidth]{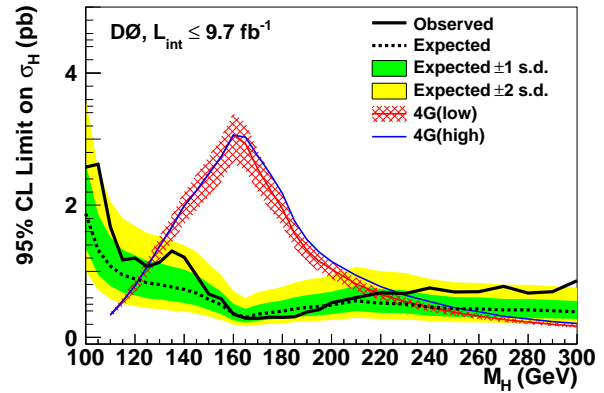}
\caption{
\label{fig:SM4} (color online)
Expected and observed 95\%\ C.L.~upper limits on Higgs boson production in fourth generation
models as a function of Higgs boson mass. The blue and red lines represent the theoretical predictions with its uncertainties
in the fourth generation ``high mass'' and ``low mass'', respectively. Below 160 GeV the models overlap and have similar
uncertainties. When setting these limits we do not include the theoretical cross section uncertainties. The shaded 
bands correspond to the regions enclosing $\pm1$ and $\pm2$~s.d.~fluctuations of the background.}
\end{figure}

\begin{figure}[htbp]
\includegraphics[width=0.45\textwidth]{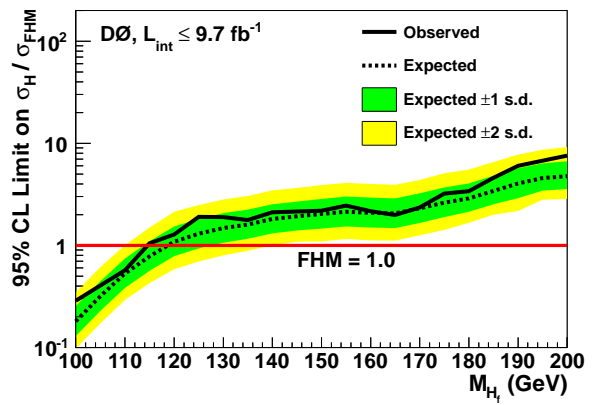}
\caption{
\label{fig:FHM} (color online)
Expected and observed 95\%\ C.L.~upper limits on fermiophobic Higgs boson production as function of Higgs boson mass. 
We exclude a fermiophobic Higgs boson with a mass below 114~GeV. The shaded bands correspond to the
regions enclosing $\pm1$ and $\pm2$~s.d.~fluctuations of the
background.}
\end{figure}

\section{Conclusions\label{sec:conclusions}}
We have presented a combination of searches for SM Higgs boson
production with the D0 experiment using data corresponding to
up to $\lumimax$~\ifb\ of $p\bar{p}$ collisions
at $\sqrt{s}=1.96$~TeV.  We set upper limits on the production
cross section at 95\%\ C.L.~for Higgs boson masses of $90 <M_H < 200$~GeV. 
We also interpret the searches in terms of models containing
a fourth generation of fermions, as well as models with a fermiophobic Higgs 
boson ($H_f$) having suppressed couplings to fermions.
We exclude a Higgs boson in the mass range $125 < M_{H} < 218$ ($125 < M_{H} < 228$)~GeV,
in the low mass (high mass) fourth generation scenario,
and a fermiophobic Higgs boson with a 
mass $100 < M_{H_f} < 114$~GeV.
The observed upper limits on SM Higgs boson production are 
$\obsABE~(\obsAFE)\times\sigma_{\text{SM}}$
at $M_H=125~(165)$~GeV, with an expected limit
of $\expABE~(\expAFE)\times\sigma_{\text{SM}}$. We exclude the 
regions of $90 < M_{H} < \excllow$~GeV and $\exclmin <M_H<
\exclmax$~GeV with an {\it a priori} expected exclusion of
$\exclminexp <M_H<\exclmaxexp$~GeV.
In the range of $M_H\approx 120-145$~GeV, the data exhibit an excess above
the background prediction of up to two standard
deviations consistent with the presence of a 125~GeV SM Higgs boson.
Each of the four main Higgs boson decay mode combinations contributes to this excess. The analyses combined here also provide inputs to the overall Tevatron combination~\cite{Aaltonen:2013kxa}, which reports an excess in data at the
level of 3 standard deviations, consistent with the production of a 
125 GeV SM Higgs boson in final states corresponding to its expected decay modes.

\begin{acknowledgments}
%
We thank the staffs at Fermilab and collaborating institutions,
and acknowledge support from the
DOE and NSF (USA);
CEA and CNRS/IN2P3 (France);
MON, NRC KI and RFBR (Russia);
CNPq, FAPERJ, FAPESP and FUNDUNESP (Brazil);
DAE and DST (India);
Colciencias (Colombia);
CONACyT (Mexico);
NRF (Korea);
FOM (The Netherlands);
STFC and the Royal Society (United Kingdom);
MSMT and GACR (Czech Republic);
BMBF and DFG (Germany);
SFI (Ireland);
The Swedish Research Council (Sweden);
and
CAS and CNSF (China).

\end{acknowledgments}

\bibliography{higgs}

\begin{thebibliography}{10}

\bibitem{Higgs:1964ia}
P.~W. Higgs,
\newblock Phys. Lett. {\bf 12}, 132 (1964).

\bibitem{Englert:1964et}
F.~Englert and R.~Brout,
\newblock Phys. Rev. Lett. {\bf 13}, 321 (1964).

\bibitem{Higgs:1964pj}
P.~W. Higgs,
\newblock Phys. Rev. Lett. {\bf 13}, 508 (1964).

\bibitem{Guralnik:1964eu}
G.~S. Guralnik, C.~R. Hagen, and T.~W.~B. Kibble,
\newblock Phys. Rev. Lett. {\bf 13}, 585 (1964).

\bibitem{Aaltonen:2012bp}
T.~Aaltonen {\em et~al.}, {[CDF Collaboration]},
\newblock Phys. Rev. Lett. {\bf 108}, 151803 (2012).

\bibitem{Abazov:2012bv}
V.~M. Abazov {\em et~al.}, [D0 Collaboration],
\newblock Phys. Rev. Lett. {\bf 108}, 151804 (2012).

\bibitem{Aaltonen:2012ra}
T.~Aaltonen {\em et~al.}, [CDF and D0 Collaboration],
\newblock Phys. Rev. D {\bf 86}, 092003 (2012).

\bibitem{bib:LEPEWWG}
{LEP Electroweak Working Group, Status as of March 2012},
\newblock {http://lepewwg.web.cern.ch/LEPEWWG/}.

\bibitem{Barate:2003sz}
R.~Barate {\em et~al.}, {(LEP Working Group for Higgs boson searches)},
\newblock Phys. Lett. B {\bf 565}, 61 (2003).

\bibitem{CDFandD0:2012aa}
TEVNPH (Tevatron New Phenomena and Higgs Working Group),
\newblock arXiv:1203.3774,
\newblock (2012).

\bibitem{Aaltonen:2010sv}
T.~Aaltonen {\em et~al.}, [CDF and D0 Collaboration],
\newblock Phys. Rev. D {\bf 82}, 011102 (2010).

\bibitem{Aad:2012an}
G.~Aad {\em et~al.}, [ATLAS Collaboration],
\newblock Phys. Rev. D {\bf 86}, 032003 (2012).

\bibitem{Chatrchyan:2012tx}
S.~Chatrchyan {\em et~al.}, [CMS Collaboration],
\newblock Phys. Lett. B {\bf 710}, 26 (2012).

\bibitem{atlas-obs}
G.~Aad {\em et~al.}, [ATLAS Collaboration],
\newblock Phys. Lett. B {\bf 716}, 1 (2012).

\bibitem{cms-obs}
S.~Chatrchyan {\em et~al.}, [CMS Collaboration],
\newblock Phys. Lett. B {\bf 716}, 30 (2012).

\bibitem{Aaltonen:2012qt}
T.~Aaltonen {\em et~al.}, [CDF and D0 Collaboration],
\newblock Phys. Rev. Lett. {\bf 109}, 071804 (2012).

\bibitem{Abachi:1993em}
S.~Abachi {\em et~al.}, {[D0 Collaboration]},
\newblock Nucl. Instrum. Methods Phys. Res. A {\bf 338}, 185 (1994).

\bibitem{Abazov:2005pn}
V.~M. Abazov {\em et~al.}, {[D0 Collaboration]},
\newblock Nucl. Instrum. Methods Phys. Res. A {\bf 565}, 463 (2006).

\bibitem{Abolins:2007yz}
M.~Abolins {\em et~al.},
\newblock Nucl. Instrum. Methods Phys. Res. A {\bf 584}, 75 (2008).

\bibitem{Angstadt:2009ie}
R.~Angstadt {\em et~al.},
\newblock Nucl. Instrum. Methods Phys. Res. A {\bf 622}, 298 (2010).

\bibitem{Abazov:2012wh97}
V.~M. Abazov {\em et~al.}, [D0 Collaboration],
\newblock Phys. Rev. Lett. {\bf 109}, 121804 (2012).

\bibitem{Abazov:lvjets}
V.~M. Abazov {\em et~al.}, [D0 Collaboration],
\newblock Phys. Rev. D {\bf 88}, 052008 (2013).

\bibitem{Abazov:2012kg}
V.~M. Abazov {\em et~al.}, [D0 Collaboration],
\newblock Phys. Rev. Lett. {\bf 109}, 121803 (2012).

\bibitem{Abazov:2013mla}
V.~M. Abazov {\em et~al.}, [D0 Collaboration],
\newblock Phys. Rev. D {\bf 88}, 052010 (2013).

\bibitem{Abazov:2012hv}
V.~M. Abazov {\em et~al.}, [D0 Collaboration],
\newblock Phys. Lett. B {\bf 716}, 285 (2012).

\bibitem{Abazov:hWWdilep}
V.~M. Abazov {\em et~al.}, [D0 Collaboration].

\bibitem{Abazov:2012zj}
V.~M. Abazov {\em et~al.}, [D0 Collaboration],
\newblock Phys. Lett. B {\bf 714}, 237 (2012).

\bibitem{Abazov:2013eha}
V.~M. Abazov {\em et~al.}, [D0 Collaboration],
\newblock Phys. Rev. D {\bf 88}, 052009 (2013).

\bibitem{Abazov:2012ee}
V.~M. Abazov {\em et~al.}, [D0 Collaboration],
\newblock Phys. Rev. D {\bf 88}, 052005 (2013).

\bibitem{Abazov:hgg}
V.~M. Abazov {\em et~al.}, [D0 Collaboration],
\newblock Phys. Rev. D {\bf 88}, 052007 (2013).

\bibitem{Abazov:2010ab}
V.~M. Abazov {\em et~al.}, {[D0 Collaboration]},
\newblock Nucl. Instrum. Methods Phys. Res. A {\bf 620}, 490 (2010).

\bibitem{Abazov:2012tf}
V.~M. Abazov {\em et~al.}, [D0 Collaboration],
\newblock Phys. Rev. Lett. {\bf 109}, 121802 (2012).

\bibitem{narsky-0507157}
I.~Narsky,
\newblock arXiv:physics/0507157,
\newblock (2005).

\bibitem{Breiman1984}
L.~Breiman, J.~H. Friedman, R.~A. Olshen, and C.~J. Stone,
\newblock {\em Classification and Regression Trees} (Wadsworth \& Brooks/Cole
  Advanced Books and Software, Pacific Grove, CA, 1984).

\bibitem{schapire01boostapproach}
R.~E. Schapire,
\newblock {T}he {B}oosting {A}pproach to {M}achine {L}earning: {A}n {O}verview,
\newblock MSRI Workshop on Nonlinear Estimation and Classification, Berkeley,
  CA, USA, 2001.

\bibitem{schapireFreund}
Y.~Freund and R.~E. Schapire,
\newblock J. Japanese Society for Artificial Intelligence {\bf 14}, 771 (1999).

\bibitem{friedman}
J.~H. Friedman,
\newblock eConf {\bf C030908}, WEAT003 (2003).

\bibitem{Hocker:2007ht}
A.~Hoecker {\em et~al.},
\newblock PoS {\bf ACAT}, 040 (2007),
\newblock we use version 4.1.0.

\bibitem{Sjostrand:2006za}
T.~Sj{\"{o}}strand, S.~Mrenna, and P.~Z. Skands,
\newblock J. High Energy Phys. {\bf 05}, 026 (2006).

\bibitem{Mangano:2002ea}
M.~L. Mangano, M.~Moretti, F.~Piccinini, R.~Pittau, and A.~D. Polosa,
\newblock J. High Energy Phys. {\bf 07}, 001 (2003).

\bibitem{Gleisberg:2008ta}
T.~Gleisberg {\em et~al.},
\newblock J. High Energy Phys. {\bf 02}, 007 (2009).

\bibitem{Boos:2004kh}
E.~Boos {\em et~al.},
\newblock Nucl. Instrum. Methods Phys. Res. A {\bf 534}, 250 (2004).

\bibitem{Boos:2006af}
E.~E. Boos, V.~E. Bunichev, L.~V. Dudko, V.~I. Savrin, and V.~V. Sherstnev,
\newblock Phys. Atom. Nucl. {\bf 69}, 1317 (2006).

\bibitem{Lai:1996mg}
H.~L. Lai {\em et~al.},
\newblock Phys. Rev. D {\bf 55}, 1280 (1997).

\bibitem{Nadolsky:2008zw}
P.~M. Nadolsky {\em et~al.},
\newblock Phys. Rev. D {\bf 78}, 013004 (2008).

\bibitem{Hamberg:1990np}
R.~Hamberg, W.~L. van Neerven, and T.~Matsuura,
\newblock Nucl. Phys. B {\bf 359}, 343 (1991),
\newblock ibid, B~\textbf{644}, 403 (2002).

\bibitem{Campbell:1999ah}
J.~M. Campbell and R.~K. Ellis,
\newblock Phys. Rev. D {\bf 60}, 113006 (1999).

\bibitem{mcfm_code}
J.~M. Campbell, R.~K. Ellis, and C.~Williams,
\newblock {MCFM - Monte Carlo for FeMtobarn processes},
\newblock \url{http://mcfm.fnal.gov/}.

\bibitem{Langenfeld:2009wd}
U.~Langenfeld, S.~Moch, and P.~Uwer,
\newblock Phys. Rev. D {\bf 80}, 054009 (2009).

\bibitem{Kidonakis:2006bu}
N.~Kidonakis,
\newblock Phys. Rev. D {\bf 74}, 114012 (2006).

\bibitem{Abazov:2007nt}
V.~M. Abazov {\em et~al.}, [D0 Collaboration],
\newblock Phys. Rev. Lett. {\bf 100}, 102002 (2008).

\bibitem{Melnikov:2006kv}
K.~Melnikov and F.~Petriello,
\newblock Phys. Rev. D {\bf 74}, 114017 (2006).

\bibitem{powheg}
T.~Melia, P.~Nason, R.~Rontsch, and G.~Zanderighi,
\newblock J. High Energy Phys. {\bf 1111}, 078 (2011).

\bibitem{Bozzi:2003jy}
G.~Bozzi, S.~Catani, D.~de~Florian, and M.~Grazzini,
\newblock Phys. Lett. B {\bf 564}, 65 (2003).

\bibitem{Bozzi:2005wk}
G.~Bozzi, S.~Catani, D.~de~Florian, and M.~Grazzini,
\newblock Nucl. Phys. B {\bf 737}, 73 (2006).

\bibitem{deFlorian:2011xf}
D.~de~Florian, G.~Ferrera, M.~Grazzini, and D.~Tommasini,
\newblock J. High Energy Phys. {\bf 11}, 064 (2011).

\bibitem{Balazs:2000sz}
C.~Balazs, J.~Huston, and I.~Puljak,
\newblock Phys. Rev. D {\bf 63}, 014021 (2001).

\bibitem{Cao:2009md}
Q.-H. Cao, C.-R. Chen, C.~Schmidt, and C.-P. Yuan,
\newblock arXiv:0909.2305,
\newblock (2009).

\bibitem{Anastasiou:2008tj}
C.~Anastasiou, R.~Boughezal, and F.~Petriello,
\newblock J. High Energy Phys. {\bf 04}, 003 (2009).

\bibitem{deFlorian:2009hc}
D.~de~Florian and M.~Grazzini,
\newblock Phys. Lett. B {\bf 674}, 291 (2009).

\bibitem{grazziniprivate}
M.~Grazzini, 2010,
\newblock {private communication}.

\bibitem{Group:2009ad}
TEVEWG (Tevatron Electroweak Working Group),
\newblock arXiv:0903.2503,
\newblock (2009).

\bibitem{Harlander:2002wh}
R.~V. Harlander and W.~B. Kilgore,
\newblock Phys. Rev. Lett. {\bf 88}, 201801 (2002).

\bibitem{Anastasiou:2002yz}
C.~Anastasiou and K.~Melnikov,
\newblock Nucl. Phys. B {\bf 646}, 220 (2002).

\bibitem{Ravindran:2003um}
V.~Ravindran, J.~Smith, and W.~L. van Neerven,
\newblock Nucl. Phys. B {\bf 665}, 325 (2003).

\bibitem{Actis200812}
S.~Actis, G.~Passarino, C.~Sturm, and S.~Uccirati,
\newblock Phys. Lett. B {\bf 670}, 12  (2008).

\bibitem{Aglietti:2006yd}
U.~Aglietti, R.~Bonciani, G.~Degrassi, and A.~Vicini,
\newblock arXiv:hep-ph/0610033,
\newblock (2006).

\bibitem{Catani:2003zt}
S.~Catani, D.~de~Florian, M.~Grazzini, and P.~Nason,
\newblock J. High Energy Phys. {\bf 07}, 028 (2003).

\bibitem{Martin:2009bu}
A.~D. Martin, W.~J. Stirling, R.~S. Thorne, and G.~Watt,
\newblock Eur. Phys. J. C {\bf 64}, 653 (2009).

\bibitem{Alekhin:2011sk}
S.~Alekhin {\em et~al.},
\newblock arXiv:1101.0536,
\newblock (2011).

\bibitem{Botje:2011sn}
M.~Botje {\em et~al.},
\newblock arXiv:1101.0538,
\newblock (2011).

\bibitem{Stewart:2011cf}
I.~W. Stewart and F.~J. Tackmann,
\newblock Phys. Rev. D {\bf 85}, 034011 (2012).

\bibitem{Anastasiou:2009bt}
C.~Anastasiou, G.~Dissertori, M.~Grazzini, F.~Stockli, and B.~R. Webber,
\newblock J. High Energy Phys. {\bf 08}, 099 (2009).

\bibitem{Campbell:2010cz}
J.~M. Campbell, R.~K. Ellis, and C.~Williams,
\newblock Phys.~Rev.~D {\bf 81}, 074023 (2010).

\bibitem{Baglio:2010um}
J.~Baglio and A.~Djouadi,
\newblock J. High Energy Phys. {\bf 10}, 064 (2010).

\bibitem{spira_prog}
M.~Spira,
\newblock Fortran codes,
\newblock \url{http://people.web.psi.ch/spira/proglist.html}.

\bibitem{Brein:2003wg}
O.~Brein, A.~Djouadi, and R.~Harlander,
\newblock Phys. Lett. B {\bf 579}, 149 (2004).

\bibitem{Ciccolini:2003jy}
M.~L. Ciccolini, S.~Dittmaier, and M.~Kramer,
\newblock Phys. Rev. D {\bf 68}, 073003 (2003).

\bibitem{Bolzoni:2011cu}
P.~Bolzoni, F.~Maltoni, S.-O. Moch, and M.~Zaro,
\newblock Phys. Rev. D {\bf 85}, 035002 (2012).

\bibitem{Djouadi:1997yw}
A.~Djouadi, J.~Kalinowski, and M.~Spira,
\newblock Comput. Phys. Commun. {\bf 108}, 56 (1998),
\newblock {We use {\sc hdecay} Version 3.53}.

\bibitem{Butterworth:2010ym}
J.~M. Butterworth {\em et~al.},
\newblock arXiv:1003.1643,
\newblock (2010).

\bibitem{Baglio:2010ae}
J.~Baglio and A.~Djouadi,
\newblock J. High Energy Phys. {\bf 03}, 055 (2011).

\bibitem{Junk:1999kv}
T.~Junk,
\newblock Nucl. Instrum. Methods Phys. Res. A {\bf 434}, 435 (1999).

\bibitem{Read:2002hq}
A.~L. Read,
\newblock J. Phys. G {\bf 28}, 2693 (2002).

\bibitem{wade_tm}
W.~Fisher,
\newblock FERMILAB-TM-2386-E  (2007).

\bibitem{Andeen:2007zc}
T.~Andeen {\em et~al.},
\newblock FERMILAB-TM-2365  (2007).

\bibitem{Holdom:2009rf}
B.~Holdom {\em et~al.},
\newblock PMC Phys. {\bf A3}, 4 (2009), 0904.4698.

\bibitem{Arik:2005ed}
E.~Arik, O.~Cakir, S.~A. Cetin, and S.~Sultansoy,
\newblock Acta Phys. Polon. B {\bf 37}, 2839 (2006).

\bibitem{Kribs:2007nz}
G.~D. Kribs, T.~Plehn, M.~Spannowsky, and T.~M.~P. Tait,
\newblock Phys. Rev. D {\bf 76}, 075016 (2007).

\bibitem{Anastasiou:2010bt}
C.~Anastasiou, R.~Boughezal, and E.~Furlan,
\newblock J. High Energy Phys. {\bf 06}, 101 (2010).

\bibitem{Aad:2011qi}
G.~Aad {\em et~al.}, [ATLAS Collaboration],
\newblock Eur. Phys. J. C {\bf 71}, 1728 (2011).

\bibitem{CMS-SM4}
S.~Chatrchyan {\em et~al.}, [CMS Collaboration],
\newblock arXiv:1302.1764,
\newblock (2013), submitted to Phys.~Lett.~B.

\bibitem{Collaboration:2012pa}
T.~Aaltonen {\em et~al.}, [CDF Collaboration],
\newblock Phys. Lett. B {\bf 717}, 173 (2012).

\bibitem{Abazov:2011ix}
V.~M. Abazov {\em et~al.}, [D0 Collaboration],
\newblock Phys. Rev. Lett. {\bf 107}, 151801 (2011).

\bibitem{Aad:2012yq}
G.~Aad {\em et~al.}, [ATLAS Collaboration],
\newblock Eur. Phys. J. C {\bf 72}, 2157 (2012).

\bibitem{CMS:2012bd}
S.~Chatrchyan {\em et~al.}, [CMS Collaboration],
\newblock J. High Energy Phys. {\bf 09}, 111 (2012).

\bibitem{Aaltonen:2013kxa}
T.~Aaltonen {\em et~al.}, [CDF and D0 Collaboration],
\newblock Phys. Rev. D {\bf 88}, 052014 (2013).

\end{thebibliography}
\bibliographystyle{h-physrev3}

\end{document}